# Numerical Investigation on Local Non-equilibrium Flows Using a Diatomic Nonlinear Constitutive Model

Zhongzheng Jiang [*], Weifang Chen [†] and Wenwen Zhao [‡]

*College of Aeronautics and Astronautics, Zhejiang University, Hangzhou, China, 310027*

and

R.S. Myong [§]

*Department of Aerospace and System Engineering and Research Center for Aircraft Parts Technology, Gyeongsang National University, Jinju, Gyeongnam 660-701, South Korea*

**The linear Navier-Stokes-Fourier (NSF) constitutive relations are capable of simulating the near-continuum flows, but fail in description of those flows which are removed far away from local equilibrium. In this paper, a diatomic nonlinear model named as nonlinear coupled constitutive relations (NCCR), derived from Eu's generalized hydrodynamics and proposed by Myong, is presented as an alternative for simulating these hypersonic gas flows with a goal of recovering NSF's solutions in continuum regime and being superior in transition regime. To guarantee stable computation, a reliable and efficient coupled algorithm is proposed for this diatomic nonlinear constitutive model. Constitutive-curve analysis is carried out in detail to compare this coupled algorithm with Myong's previous algorithm. Local flow regions are investigated carefully in these hypersonic flows past a cone tip, a hollow cylinder-flare and a HTV-type vehicle. The convergent solutions of NCCR model are compared with NSF, DSMC calculations and experiment. It is demonstrated that the NCCR model works as efficiently as the NSF model in continuum regime, but more accurately compared with DSMC and experiment than NSF in non-equilibrium flows. The discrepancies of flow-field and surface parameters, imply a potential for remedying NSF's deficiency in local non-equilibrium regions.**

## Nomenclature

| | | |
|---|---|---|
| $C$ | = | peculiar velocity ( $m/s$ ) |
| $f$ | = | phase density |
| $k_B$ | = | Boltzmann constant |
| $\mu$ | = | normalization factor in Eu's phase density |
| $d_r$ | = | reference molecular diameter |
| $c$ | = | NCCR constant $\left(mk_BT_r\right)^{1/4}\big/2d_r\eta_r^{1/2}$ |
| $m$ | = | molecular mass |
| $E$ | = | total energy per unit mass ( $J/kg$ ) |
| $\hat{\mathbf{F}}_c, \hat{\mathbf{G}}_c, \hat{\mathbf{H}}_c$ | = | convective flux vector |
| $\hat{\mathbf{F}}_v, \hat{\mathbf{G}}_v, \hat{\mathbf{H}}_v$ | = | viscous flux vector |
| $\mathbf{A}^{\pm}, \mathbf{B}^{\pm}, \mathbf{C}^{\pm}$ | = | inviscid approximate Jacobian matrixes |
| $\rho(\mathbf{A}), \rho(\mathbf{B}), \rho(\mathbf{C})$ | = | spectral radiuses for the inviscid Jacobian matrixes. |

*Ph.D. candidate, College of Aeronautics and Astronautics, Zhejiang University, Hangzhou, China, 310027; Student member, AIAA

†Professor, College of Aeronautics and Astronautics, Zhejiang University, Hangzhou, China, 310027

‡Assistant Professor, College of Aeronautics and Astronautics, Zhejiang University, Hangzhou, China, 310027

§Professor, Department of Aerospace and System Engineering and Research Center for Aircraft Parts Technology, Gyeongsang National University, Jinju, Gyeongnam 660-701, South Korea

$\hat{\mathbf{Q}}$ = vector of flow variables in physical domain

$J$ = Jacobian coordinate

$Kn_{BLG}$ = the body-length global (BLG) Knudsen number

$L$ = characteristic length ( $m$ )

$L_0$ = reference length ( $m$ )

$l_\infty$ = mean free path of gas molecules ( $m$ )

$Ma$ = Mach number

$Re$ = Reynolds number

$N_\delta$ = ratio of Mach number to Reynolds number

Pr = Prandtl number

$p$ = pressure ( $N/m^2$ )

$p_\infty$ = free stream pressure ( $N/m^2$ )

$T$ = temperature ( $K$ )

$T_\infty$ = free stream temperature ( $K$ )

$T_{ref}$ = reference temperature ( $K$ )

$T_w$ = wall temperature

$\alpha$ = attack angle

$U_\infty$ = free stream velocity ( $m/s$ )

$u_i$ = velocity components ( $m/s$ )

$\rho$ = density ( $kg/m^3$ )

$\rho_\infty$ = free stream density ( $kg/m^3$ )

$\rho_{ref}$ = reference density ( $kg/m^3$ )

$\eta$ = viscosity ( $(N\cdot s)/m^2$ )

$\eta_{ref}$ = reference viscosity ( $(N\cdot s)/m^2$ )

$\eta_b$ = bulk viscosity ( $(N\cdot s)/m^2$ )

$\lambda$ = thermal conductivity ( $J/(m\cdot s\cdot K)$ )

$f_b$ = ratio of the bulk viscosity to the shear viscosity

$s_{VHS}$ = VHS temperature exponent

$R$ = gas constant ( $J/(kg\cdot K)$ )

$\gamma$ = specific heat ratio

$c_p$ = constant-pressure specific heat( $J/(kg\cdot K)$ )

$C_p$ = pressure coefficient

$C_f$ = skin friction coefficient

$C_h$ = heat transfer rate

$\mathbf{\Pi}$ = shear stress tensor ( $N/m^2$ )

$\Delta$ = excess normal stress ( $N/m^2$ )

$\mathbf{Q}$ = heat flux ( w/ $m^2$ )

$\psi$ = higher-order moments

$\kappa$ = first-order cumulant of cumulant approximation for dissipation terms

$q(\kappa)$ = nonlinear dissipation factor

$\hat{R}$ = dimensionless Rayleigh-Onsager dissipation function

$[\mathbf{A}]^{(2)}$ = traceless symmetric part of the second-rank tensor $\mathbf{A}$ , equal to $\left(A+A^T\right)/2-\mathbf{I}Tr\mathbf{A}/3$

$\mathbf{I}$ = identity matrix

## I. Introduction

Near space defined as the airspace between 20 to 100 kilometres high above sea level, has been investigated quite intensively in recent years. In near space, the physical properties of gas including density, pressure and temperature, vary as a function of altitude dramatically. Re-entry vehicles such as HTV (Hypersonic Technology Vehicle), space shuttle or Apollo capsule, will experience different flow regimes from rarefaction to continuum when passing through near space during the course of their flight trajectory. The local flow field generated by hypersonic vehicles at high altitudes contains typical rarefied non-equilibrium characteristics, which occur obviously in the shock waves around the sharp leading edge, Knudsen layer near the solid surface, as well as the gaseous expansion region in the near-wake of vehicles. The reproduction of such non-equilibrium phenomenon in ground-based facilities or flight tests are extremely challenging and expensive and thus numerical simulation should be a better choice for the design of such vehicles. However, as the rarefied non-equilibrium effect is strengthened, the linear laws of Navier-Stokes-Fourier (NSF) become inaccurate and are not available any more. And hence a more refined set of theoretical tools beyond the classical theory of linear constitutive relations needs to be developed.

For many years, much effort has been undertaken on getting the real physical solution of the challenging flow problems of which NSF equations fail in description. Among them, the direct simulation Monte Carlo (DSMC) method [1] is rather successful one to solve the Boltzmann equation by counting the real molecule collisions with collision frequencies and scattering velocity distributions. Other well-known methods also include linearized methods of Boltzmann equation [2], Boltzmann model equations (BGK, ES-BGK and Shakhov models) [3-5], discrete velocity methods (DVM) [6], unified gas-kinetic scheme(UGKS) [7], Burnett-type equations [8-10], Grad's moment equations [11], regularized 13 moment equations [12] etc. For prediction of aerodynamic characteristics of re-entry vehicles, it is recommended that any new methods should obey three acknowledged standards: (1) recovering the Navier-Stokes-Fourier solution in continuum regime and being superior in transition regime; (2) acceptable computational efficiency with less cost; (3) stable and accurate ability to handle complex configurations. However, either huge computational consumption or instability when handling complex geometries makes the aforementioned methods less popular in engineering application.

A set of generalized hydrodynamic equations proposed by Eu [13] provides an alternative solution procedure to the Boltzmann equation. However, Eu's 13 moment equations also have difficulty in handling multidimensional problems. In order to provide a practical high-order fluid dynamic model with stable computational capacity, Myong developed an efficient computational model on the basis of Eu's equations [14, 15]. The model, named as the nonlinear coupled constitutive relations (NCCR), takes a form of nonlinear algebraic system and can be implemented more easily in the hyperbolic conservation laws. The second-order molecular dynamic characterization of NCCR model was validated in Poiseuille flow research through the deterministic microscopic MD method [16]. Myong also constructed an uncoupled computational algorithm for this model, which was applied successfully in one-dimensional shock wave structure and two-dimensional simple flow [14, 17]. However, the uncoupled solution to NCCR model encounters numerical instability especially under three-dimensional conditions due to overlooking some significant effects among these multi-dimensional non-conserved variables [18, 19]. In present work, we attempt to utilize a coupled algorithm which combined Fixed-point and Newton's iterations for complete computation of non-conserved variables in NCCR model and modern CFD schemes to solve the conservational equations for conserved variables within finite volume framework. Satisfying and convergent NCCR results of the three-dimensional hypersonic flows are obtained and investigated carefully. Section II and III will give a briefly introduction about this model and the numerical solution process. The performance of the coupled and uncoupled algorithms is also investigated carefully in section III. Section IV will then discuss the flow field and surface property results calculated by this model and present a detailed comparison with linear NSF, DSMC and experiment prediction. Finally, some conclusions are drawn.

## II. The Diatomic Nonlinear Constitutive Model

In order to solve the far-from-equilibrium flow problems, Eu [20, 21] came up with generalized hydrodynamic equations (GHE) consistent with the laws of irreversible thermodynamics. Different from Chapman-Enskog expansion with Knudsen number as a small parameter, GHE can be derived from Boltzmann equation by ingenious construction of non-equilibrium canonical distribution function (1) and employing the cumulant-expansion method for Boltzmann collision term. The non-equilibrium canonical distribution function is defined as

$$f = \exp\left[-\frac{1}{k_B T}\left(\frac{1}{2}mC^2 + H_{rot} + \sum_{k=1}^{\infty} X_k h^{(k)} - \mu\right)\right], \tag{1}$$

where $\mu$ is the normalization factor. $X_k$ are functions of macroscopic variables and occupy the status similar to the coefficients of Maxwell-Grad moment method. $T$, $k_B$, $m$ and $H_{rot}$ represent temperature, Boltzmann constant, molecular mass and rotational Hamiltonian of molecule respectively. Finally, a set of evolution equations of non-conserved variables for a diatomic gas (GHE) can be obtained as

$$\begin{aligned}
&\rho\frac{D(\mathbf{\Pi}/\rho)}{Dt} + \nabla\cdot\psi_4 = -2[\mathbf{\Pi}\cdot\nabla\mathbf{u}]^{(2)} - \frac{p}{\eta}\mathbf{\Pi}q(\kappa) - 2(p+\Delta)[\nabla\mathbf{u}]^{(2)},\\
&\rho\frac{D(\Delta/\rho)}{Dt} + \nabla\cdot\psi_5 = -2\gamma'(\Delta\mathbf{I}+\mathbf{\Pi}):\nabla\mathbf{u} - \frac{2}{3}\gamma' p\nabla\cdot\mathbf{u} - \frac{2}{3}\gamma'\frac{p}{\eta_b}\Delta q(\kappa),\\
&\rho\frac{D(\mathbf{Q}/\rho)}{Dt} + \nabla\cdot\psi_6 = -\mathbf{\Pi}\cdot c_p\nabla T - \mathbf{Q}\cdot\nabla\mathbf{u} - \frac{pc_p}{\lambda}\mathbf{Q}q(\kappa) - (p+\Delta)c_pT\nabla\ln T + \nabla\cdot[(p+\Delta)\mathbf{I}+\mathbf{\Pi}]\cdot\frac{(\mathbf{\Pi}+\Delta\mathbf{I})}{\rho},
\end{aligned} \tag{2}$$

where $\rho$, $p$, $T$, $\mathbf{\Pi}$, $\Delta$ and $\mathbf{Q}$ denote density, hydrostatic pressure, temperature, the shear stress, the excess normal stress and the heat flux respectively. $\gamma'$ is equal to $(5-3\gamma)/2$ and the nonlinear dissipation factor $q(\kappa) = \sinh(\kappa)/\kappa$ with $\kappa = (mk_BT)^{1/4}/\sqrt{2}dp\left[\mathbf{\Pi}:\mathbf{\Pi}/2\eta + \gamma'\Delta^2/\eta_b + \mathbf{Q}\cdot\mathbf{Q}/\lambda T\right]^{1/2}$. $\psi_4$, $\psi_5$ and $\psi_6$ represent the flux of high-order moments. In order to close aforementioned evolution equations, Eu [21, 22] provided a closure (3) different from Grad's, which was considered by Myong [23] to achieve a balance between the kinematic and collision term approximation

$$\psi_4 = \psi_5 = \psi_6 = 0. \tag{3}$$

However, with the existence of time term of non-conserved variables in GHE, the numerical computation has to consider the mathematical properties of partial differential equations and their proper numerical schemes, which may bring great computational difficulties in such a nonlinear system. In order to omit the substantial time derivative terms $D(\mathbf{\Pi}/\rho)/Dt$, $D(\Delta/\rho)/Dt$ and $D(\mathbf{Q}/\rho)/Dt$ in the left hand side of GHE (2), Eu [21] proposed an adiabatic approximation assumption: the transportation of conserved variables and the non-conserved variables vary on two different time scales. The non-conserved variables change faster than the conserved variables and thus reach a steady state more quickly than the latter. On the time scale of conserved variables, the time terms for non-conserved variables can be neglected. Note that the derivative term of non-conserved variables $\nabla\cdot[(p+\Delta)\mathbf{I}+\mathbf{\Pi}]\cdot(\mathbf{\Pi}+\Delta\mathbf{I})/\rho$ is removed by Myong in order to establish an algebraic system which can be solved by iterative methods. The term $\mathbf{Q}\cdot\nabla\mathbf{u}$ is also omitted just for the sake of simplicity [15, 17]. Finally, a nonlinear algebraic system of the second-order nonlinear model, namely the nonlinear coupled constitutive relations (NCCR), is developed. In our recent work [24], the insignificant effect from this term $\mathbf{Q}\cdot\nabla\mathbf{u}$ has also been validated. The diatomic NCCR model can be summarized below as

$$
\begin{aligned}
&-2\left[\mathbf{\Pi}\cdot\nabla\mathbf{u}\right]^{(2)}-\frac{p}{\eta}\mathbf{\Pi}q\left(\kappa\right)-2(p+\Delta)\left[\nabla\mathbf{u}\right]^{(2)}=0,\\
&-2\gamma'(\Delta\mathbf{I}+\mathbf{\Pi}):\nabla\mathbf{u}-\frac{2}{3}\gamma' p\nabla\cdot\mathbf{u}-\frac{2}{3}\gamma'\frac{p}{\eta_b}\Delta q\left(\kappa\right)=0,\\
&-\mathbf{\Pi}\cdot c_p\nabla T-\frac{pc_p}{\lambda}\mathbf{Q}q\left(\kappa\right)-(p+\Delta)c_pT\nabla\ln T=0.
\end{aligned}
\tag{4}
$$

The linear Navier-Stokes-Fourier constitutive relations, derived from the first-order Chapman-Enskog expansion, can be summarized as a comparison with the second-order model (4),

$$
\mathbf{\Pi}_0=-2\eta\left[\nabla\mathbf{u}\right]^{(2)},\quad \Delta_0=-\eta_b\nabla\cdot\mathbf{u},\quad \mathbf{Q}_0=-\lambda\nabla T.
\tag{5}
$$

The symbol $\left[\mathbf{A}\right]^{(2)}$ represents the traceless symmetric part of the second-rank tensor $\mathbf{A}$. It is worthwhile mentioning that the bulk viscosity is equal to zero in the linear theory according to Stokes's hypothesis. Introducing NSF relations (5) into NCCR model (4) and nondimensionalizing with infinite flow parameters [18], we can get another more concise form of NCCR model as

$$
\begin{aligned}
&\hat{\mathbf{\Pi}}q\left(c\hat{R}\right)=(1+\hat{\Delta})\hat{\mathbf{\Pi}}_0+\left[\hat{\mathbf{\Pi}}\cdot\nabla\hat{\mathbf{u}}\right]^{(2)},\\
&\hat{\Delta}q\left(c\hat{R}\right)=\hat{\Delta}_0+\frac{3}{2}f_b(\hat{\Delta}\mathbf{I}+\hat{\mathbf{\Pi}}):\nabla\hat{\mathbf{u}},\\
&\hat{\mathbf{Q}}q\left(c\hat{R}\right)=(1+\hat{\Delta})\hat{\mathbf{Q}}_0+\hat{\mathbf{\Pi}}\cdot\hat{\mathbf{Q}}_0,
\end{aligned}
\tag{6}
$$

where

$$
\begin{aligned}
&\frac{\eta_\infty}{\rho_\infty a_\infty L_0}=N_\delta=\frac{\mathrm{Ma}}{\mathrm{Re}},\quad \hat{\mathbf{\Pi}}=\frac{N_\delta}{p}\mathbf{\Pi},\quad \hat{\Delta}=\frac{N_\delta}{p}\Delta,\quad \hat{\mathbf{Q}}=\frac{N_\delta}{p}\frac{\mathbf{Q}}{\sqrt{T/(2\varepsilon)}}\\
&\hat{R}=\left[\hat{\mathbf{\Pi}}:\hat{\mathbf{\Pi}}+\frac{2\gamma'}{f_b}\hat{\Delta}^2+\hat{\mathbf{Q}}\cdot\hat{\mathbf{Q}}\right]^{1/2},\nabla\hat{\mathbf{u}}=-2\eta\frac{N_\delta}{p}\nabla\mathbf{u},\varepsilon=\frac{1}{\Pr(\gamma-1)}
\end{aligned}
$$

In eqs.(6), the constant $c$ is defined as $\left(mk_BT_r\right)^{1/4}/2d_r\eta_r^{1/2}$.

## III. Numerical Solution

### A. Governing equation and temporal-spatial discretization

The three-dimensional governing equations of conserved variables for a diatomic gas in conservation law form in curvilinear coordinate system can be expressed as:

$$
\frac{\partial\hat{\mathbf{Q}}}{\partial\tau}+\frac{\partial\hat{\mathbf{F}}_c}{\partial\xi}+\frac{\partial\hat{\mathbf{G}}_c}{\partial\eta}+\frac{\partial\hat{\mathbf{H}}_c}{\partial\zeta}+N_\delta\left(\frac{\partial\hat{\mathbf{F}}_v}{\partial\xi}+\frac{\partial\hat{\mathbf{G}}_v}{\partial\eta}+\frac{\partial\hat{\mathbf{H}}_v}{\partial\zeta}\right)=0,
\tag{7}
$$

where

$$\hat{\mathbf{Q}} = \frac{1}{J}\begin{bmatrix} \rho \\ \rho u \\ \rho v \\ \rho w \\ \rho E \end{bmatrix},\ \hat{\mathbf{F}}_c = \frac{1}{J}\begin{bmatrix} \rho U \\ \rho uU + p\xi_x \\ \rho vU + p\xi_y \\ \rho wU + p\xi_z \\ (\rho E + p)U - p\xi_t \end{bmatrix},\ \hat{\mathbf{G}}_c = \frac{1}{J}\begin{bmatrix} \rho V \\ \rho uV + p\eta_x \\ \rho vV + p\eta_y \\ \rho wV + p\eta_z \\ (\rho E + p)V - p\eta_t \end{bmatrix}$$

$$,\ \hat{\mathbf{H}}_c = \frac{1}{J}\begin{bmatrix} \rho W \\ \rho uW + p\zeta_x \\ \rho vW + p\zeta_y \\ \rho wW + p\zeta_z \\ (\rho E + p)W - p\zeta_t \end{bmatrix},\hat{\mathbf{F}}_v = \frac{1}{J}\begin{bmatrix} 0 \\ \xi_x\left(\Pi_{xx}+\Delta\right)+\xi_y\Pi_{xy}+\xi_z\Pi_{xz} \\ \xi_x\Pi_{xy}+\xi_y\left(\Pi_{yy}+\Delta\right)+\xi_z\Pi_{yz} \\ \xi_x\Pi_{xz}+\xi_y\Pi_{yz}+\xi_z\left(\Pi_{zz}+\Delta\right) \\ \xi_x b_x+\xi_y b_y+\xi_z b_z \end{bmatrix},$$

$$\hat{\mathbf{G}}_v = \frac{1}{J}\begin{bmatrix} 0 \\ \eta_x\left(\Pi_{xx}+\Delta\right)+\eta_y\Pi_{xy}+\eta_z\Pi_{xz} \\ \eta_x\Pi_{xy}+\eta_y\left(\Pi_{yy}+\Delta\right)+\eta_z\Pi_{yz} \\ \eta_x\Pi_{xz}+\eta_y\Pi_{yz}+\eta_z\left(\Pi_{zz}+\Delta\right) \\ \eta_x b_x+\eta_y b_y+\eta_z b_z \end{bmatrix},\hat{\mathbf{H}}_v = \frac{1}{J}\begin{bmatrix} 0 \\ \zeta_x\left(\Pi_{xx}+\Delta\right)+\zeta_y\Pi_{xy}+\zeta_z\Pi_{xz} \\ \zeta_x\Pi_{xy}+\zeta_y\left(\Pi_{yy}+\Delta\right)+\zeta_z\Pi_{yz} \\ \zeta_x\Pi_{xz}+\zeta_y\Pi_{yz}+\zeta_z\left(\Pi_{zz}+\Delta\right) \\ \zeta_x b_x+\zeta_y b_y+\zeta_z b_z \end{bmatrix}. \tag{8}$$

In formulas (8), $J$ is the Jacobian coordinate transformation and $U, V, W$ are contravariant velocities, expressed as

$$\begin{aligned} U &= \xi_x u + \xi_y v + \xi_z w + \xi_t, \\ V &= \eta_x u + \eta_y v + \eta_z w + \eta_t, \\ W &= \zeta_x u + \zeta_y v + \zeta_z w + \zeta_t. \end{aligned}$$

and

$$\begin{aligned} b_x &= u\left(\Pi_{xx}+\Delta\right)+v\Pi_{xy}+w\Pi_{xz}+\varepsilon q_x, \\ b_y &= u\Pi_{yx}+v\left(\Pi_{yy}+\Delta\right)+w\Pi_{yz}+\varepsilon q_y, \\ b_z &= u\Pi_{zx}+v\Pi_{zy}+w\left(\Pi_{zz}+\Delta\right)+\varepsilon q_z. \end{aligned}$$

The linear Navier-Stokes-Fourier constitutive relations can be rewritten in an index form as

$$\begin{aligned} \Pi_{ij0} &= -\eta\left(\frac{\partial u_j}{\partial x_i}+\frac{\partial u_i}{\partial x_j}\right)+\frac{2\eta}{3}\frac{\partial u_k}{\partial x_k}\delta_{ij}, \\ \Delta_0 &= -\eta_b\frac{\partial u_i}{\partial x_i} = 0, \\ Q_{i0} &= -\lambda\frac{\partial T}{\partial x_i}. \end{aligned} \tag{9}$$

The linear constitutive relations (9) in conjunction with Eqs (7) and (8) yield the well-known NS equations. In contrast, the dimensionless nonlinear coupled constitutive relations are written as

$$\begin{aligned} q\left(c\hat{R}\right)\hat{\Pi}_{ij} &= \left(1+\hat{\Delta}\right)\hat{\Pi}_{ij0}+\frac{1}{2}\left(\hat{\Pi}_{ik}\frac{\partial\hat{u}_j}{\partial x_k}+\hat{\Pi}_{jk}\frac{\partial\hat{u}_i}{\partial x_k}\right)-\frac{1}{3}\hat{\Pi}_{mk}\frac{\partial\hat{u}_m}{\partial x_k}\delta_{ij}, \\ q\left(c\hat{R}\right)\hat{\Delta} &= \hat{\Delta}_0+\frac{3}{2}f_b\left(\hat{\Pi}_{ij}+\hat{\Delta}\delta_{ij}\right)\frac{\partial\hat{u}_j}{\partial x_i}, \\ q\left(c\hat{R}\right)\hat{Q}_i &= \left(1+\hat{\Delta}\right)\hat{Q}_{i0}+\hat{\Pi}_{ij}\hat{Q}_{j0}, \end{aligned} \tag{10}$$

where $\hat{R}^2 = \hat{\Pi}_{ij}^2 + (2\gamma'/f_b)\hat{\Delta}^2 + \hat{Q}_i^2$. Together with Eqs. (7) and (8), the nonlinear relations (10) yield so-called NCCR equations.

The NCCR equations are attempted to be solved numerically using the finite volume method. For computing the inviscid flux of hyperbolic conservational system, AUSMPW+ scheme was proposed by Kim [25] through introducing the pressure-based weight function to remove the oscillations of AUSM+ and overcome carbuncle phenomena of AUSMD. This flux splitting scheme is extensively used in hypersonic flows due to its efficiency, robustness and strong capability in capturing the shock. Therefore, the AUSMPW+ is also employed in computing the convective flux of NCCR equations. The AUSMPW+ scheme can be summarized as

$$F_{1/2} = c_{1/2}\left(\bar{M}_L^+\Phi_L + \bar{M}_R^-\Phi_R\right) + \left(P_L^+\Big|_{\alpha=3/16}\mathbf{P}_L + P_R^-\Big|_{\alpha=3/16}\mathbf{P}_R\right). \tag{11}$$

For the detail of each parameter in (11) see the literature [25]. Note that the description of gas viscosity is significant in simulating rarefied flows. High-resolution Van Albada limiter with less dissipation is utilized together with Van Leer's MUSCL (Monotonic Upstream-Centred Scheme for Conservation Laws) approach to reconstruct the left and right values of the interface, which is summarized as

$$\begin{aligned} q_{i+1/2}^L &= q_i + \frac{1}{4}\left[(1-\kappa)\bar{\Delta}_- + (1+\kappa)\bar{\Delta}_+\right]_i, \\ q_{i+1/2}^R &= q_{i+1} - \frac{1}{4}\left[(1-\kappa)\bar{\Delta}_+ + (1+\kappa)\bar{\Delta}_-\right]_{i+1}, \end{aligned} \tag{12}$$

where

$$\begin{aligned} &\bar{\Delta}_- = \bar{\Delta}_+ = Van\ albada(\Delta_-, \Delta_+), \\ &(\Delta_-)_i = q_i - q_{i-1},\ (\Delta_+)_i = q_{i+1} - q_i, \\ &Van\ albada(x,y) = \frac{x(y^2+\varepsilon) + y(x^2+\varepsilon)}{x^2+y^2+2\varepsilon}. \end{aligned}$$

The spatial derivatives of viscous flux are discretized by the second-order central differencing. For the time discretization, the implicit Lower-Upper Symmetric Gauss-Seidel (LU-SGS) scheme is adopted to get an efficient steady convergence. The governing equation (7) by the first-order implicit temporal discretization can be rewritten as

$$\left[\mathbf{I} + \Delta t\left(D_\xi\mathbf{A} + D_\eta\mathbf{B} + D_\zeta\mathbf{C}\right)\right]\Delta Q^n = -\Delta t\mathbf{RHS}, \tag{13}$$

where $D_\xi$, $D_\eta$, $D_\zeta$ are differential operators and $\mathbf{A}$, $\mathbf{B}$, $\mathbf{C}$ are inviscid Jacobian matrixes. $\mathbf{RHS}$ is residual. Finally the LUSGS time-marching form can be got by LU split as

$$\mathbf{L}\mathbf{D}^{-1}\mathbf{U}\Delta Q = -\mathbf{RHS}, \tag{14}$$

where

$$\begin{aligned} \mathbf{D} &= \mathbf{I}\left[\frac{1}{\Delta t} + \rho(\mathbf{A}) + \rho(\mathbf{B}) + \rho(\mathbf{C})\right], \\ \mathbf{L} &= \mathbf{D} - \mathbf{A}_{i-1}^+ - \mathbf{B}_{j-1}^+ - \mathbf{C}_{k-1}^+, \\ \mathbf{U} &= \mathbf{D} + \mathbf{A}_{i+1}^- + \mathbf{B}_{j+1}^- + \mathbf{C}_{k+1}^-. \end{aligned}$$

$\mathbf{A}^\pm$, $\mathbf{B}^\pm$, $\mathbf{C}^\pm$ are approximate Jacobian matrixes and $\rho(\mathbf{A})$, $\rho(\mathbf{B})$, $\rho(\mathbf{C})$ are spectral radiuses for the inviscid Jacobian matrixes. In general, explicit temporal discretization of viscous term in (7) would lead to computational instability in viscosity-predominated region, such as boundary layer. In this work, we try to approximately discretize the viscous term in implicit way by utilizing the viscous spectral radiuses to modify the eigenvalues for approximate Jacobian matrixes. Take $\xi$ direction for instance,

$$\bar{A}^{\pm} = A^{\pm} + kI \tag{15}$$

Note that the Jacobian matrixes for NCCR viscous terms are extremely complex to deduce the viscous spectral radiuses directly. The NSF's viscous spectral radiuses $k = 2Ma_{\infty}\mu|\nabla\xi|^2 / \mathrm{Re}\,\rho$ are utilized for NCCR equations. Overall, the computational results show that this approximation meets the requirement of efficiency and stability for NCCR equations.

**B. Coupled algorithm and constitutive-curve analysis**

The only difference between NS and NCCR equations is the solution process of viscous stresses and heat flux. They can be computed directly and explicitly from the first-order derivative of conserved variable in NSF relations (9) but indirectly and implicitly in NCCR model (10). Note that there is no derivative of these non-conserved variables $\hat{\Pi}_{ij}$, $\hat{\Delta}$, $\hat{Q}_i$ in NCCR model (10). Therefore, an additional iterative process is available by firstly modelling NCCR model as a general form of nonlinear algebraic equation systems as below:

$$\begin{aligned}
&f_1\left(\hat{\Pi}_{xx},\hat{\Pi}_{yy},\hat{\Pi}_{xy},\hat{\Pi}_{xz},\hat{\Pi}_{yz},\hat{Q}_x,\hat{Q}_y,\hat{Q}_z,\hat{\Delta}\right)=0,\\
&f_2\left(\hat{\Pi}_{xx},\hat{\Pi}_{yy},\hat{\Pi}_{xy},\hat{\Pi}_{xz},\hat{\Pi}_{yz},\hat{Q}_x,\hat{Q}_y,\hat{Q}_z,\hat{\Delta}\right)=0,\\
&\qquad\vdots\\
&f_9\left(\hat{\Pi}_{xx},\hat{\Pi}_{yy},\hat{\Pi}_{xy},\hat{\Pi}_{xz},\hat{\Pi}_{yz},\hat{Q}_x,\hat{Q}_y,\hat{Q}_z,\hat{\Delta}\right)=0,
\end{aligned} \tag{16}$$

where $f_i$ can be defined as a function of 9 no-derivative independent variables $\left(\hat{\Pi}_{xx},\hat{\Pi}_{yy},\hat{\Pi}_{xy},\hat{\Pi}_{xz},\hat{\Pi}_{yz},\hat{Q}_x,\hat{Q}_y,\hat{Q}_z,\hat{\Delta}\right)$, mapping a 9-dimensional space $\mathbb{R}^9$ into the real line $\mathbb{R}$. A more compact vector form of the nonlinear equation systems (16) can be rewritten as:

$$\mathbf{F}(\mathbf{x}) = 0, \tag{17}$$

where $\mathbf{F} = (f_1, f_2, \ldots, f_9)^t$ and $\mathbf{x}=\left(\hat{\Pi}_{xx},\hat{\Pi}_{yy},\hat{\Pi}_{xy},\hat{\Pi}_{xz},\hat{\Pi}_{yz},\hat{Q}_x,\hat{Q}_y,\hat{Q}_z,\hat{\Delta}\right)^t$

Two numerical approximation methods, the Fixed-point iteration $\mathbf{x} = G(\mathbf{x})$ and Newton's method $\mathbf{G}(\mathbf{x}) = \mathbf{x} - J(\mathbf{x})^{-1}\mathbf{F}(\mathbf{x})$ , are attempted to be used for solving aforementioned systems of nonlinear algebraic equations (17). In Fixed-point iteration, the first challenge is to construct an available convergent iterative expression for NCCR model. We intend to merges evolution equations (10) for NCCR's non-conserved variables (shear stress, bulk stresses and heat flux) into one formulation $q\left(c\hat{R}\right)\hat{R}^2 = F$ using the Rayleigh-Onsager dissipation function, in the sense that it conserves the whole information of non-conserved variables by one scalar formulation

$$\begin{aligned}
F = {}& \hat{\Pi}_{ij}\left[\left(1+\hat{\Delta}\right)\hat{\Pi}_{ij0} + \frac{1}{2}\left(\hat{\Pi}_{ik}\frac{\partial \hat{u}_j}{\partial x_k} + \hat{\Pi}_{jk}\frac{\partial \hat{u}_i}{\partial x_k}\right) - \frac{1}{3}\hat{\Pi}_{mk}\frac{\partial \hat{u}_m}{\partial x_k}\delta_{ij}\right] \\
& + (2\gamma'/f_b)\hat{\Delta}\left[\hat{\Delta}_0 + \frac{3}{2}f_b\left(\hat{\Pi}_{ij} + \hat{\Delta}\delta_{ij}\right)\frac{\partial \hat{u}_j}{\partial x_i}\right] + \left[\left(1+\hat{\Delta}\right)\hat{Q}_{i0} + \hat{\Pi}_{ij}\hat{Q}_{j0}\right]\hat{Q}_i,
\end{aligned}$$

and improves the computation efficient by reducing the amount of iterative scalar equations. $\hat{R}$ serves as an intermediate variable in the iteration $\hat{R}_{n+1} = \sinh^{-1}\left(cF_n / \hat{R}_n\right) / c$. The Fixed-point iterative formulation is stable, but unfortunately can merely cover most domains rather than the whole range in $\hat{u}_x$-only and $\hat{v}_x$-only testing problems. On the other hand, although Newton's method is generally expected to have quadratic convergence, it extremely depends on a sufficiently accurate starting value and invertibility of matrix $J(\mathbf{x})^{-1}$. Owing to NCCR model's complex mathematical properties and subject to single method's iterative limitation, it is hard to obtain convergence for all flow regimes incorporating a wide range of local gradients of flow parameters by a single method. But an interesting finding shows that the combination of these two iterative methods' convergent regions together could cover the whole range of flow gradients for NCCR model. Therefore, a coupled algorithm with combination of Fixed-point and Newton's iterations is proposed for solving the NCCR model on the basis of their convergent conditions. In our coupled solver, we switch to use one of the two iterative methods in the computation of the flow region when the other one fails to yield convergent solution.

The diatomic nonlinear constitutive curves in one-dimensional compression-expansion ($\hat{u}_x$-only) and shear ($\hat{v}_x$-only) problems are analysed to compare our coupled solution and Myong's previous works [14, 15, 17]. In these two simple flow problems we try to demonstrate the variation of each stress component with the derivative of velocities $\hat{u}$ and $\hat{v}$ in x coordinate direction. Figure 1 shows that the normal stress components $\hat{\Pi}_{xx}$, $\hat{\Pi}_{yy}$, $\hat{\Pi}_{zz}$ and the excess normal stress $\hat{\Delta}$ vary with $\hat{u}_x$, while Figure 2 demonstrates their relations with $\hat{v}_x$. Note that the normal stresses $\hat{\Pi}_{yy}$ and $\hat{\Pi}_{zz}$ cannot be computed directly in two solvers (one for $\hat{u}_x$-only and another for $\hat{v}_x$-only problems) of the uncoupled algorithm. Therefore, the uncoupled solutions of $\hat{\Pi}_{yy}$ and $\hat{\Pi}_{zz}$ originate from two simple relations, $\hat{\Pi}_{yy}=\hat{\Pi}_{zz}=-\hat{\Pi}_{xx}/2$ for $\hat{u}_x$–only and $\hat{\Pi}_{xx}=\hat{\Pi}_{zz}=-\hat{\Pi}_{yy}/2$ for $\hat{v}_x$ -only. To some extent $\hat{u}_x$ and $\hat{v}_x$ can be regarded as the level of departure from the local equilibrium state. All the NCCR figures demonstrate strong nonlinearity in comparison with the linear NSF constitutive relations. NCCR model yields a great discrepancy when $\hat{u}_x, \hat{v}_x \to \infty, -\infty$, but could recover linear NSF solutions in the local equilibrium $\hat{u}_x = \hat{v}_x = 0$, which implies a physical potential to supplement the conventional NSF model. Furthermore, we can see that coupled solutions yield excellent agreement with uncoupled solutions of these normal stresses $\hat{\Pi}_{xx}$, $\hat{\Pi}_{yy}$, $\hat{\Pi}_{zz}$, shear stress $\hat{\Pi}_{xy}$ and excess normal stress $\hat{\Delta}$. It means that our coupled NCCR solver developed for multi-dimensional problems can recover one-dimensional solutions in simple flow problems.

In order to highlight the improvement in solving the NCCR model in comparison with the uncoupled algorithm, a coupled compression-expansion-shear flow is considered by combining the $\hat{u}_x$– and $\hat{v}_x$–induced multi-dimensional effects. Figure 3 demonstrates the discrepancy between the coupled and uncoupled solutions of $\hat{\Pi}_{xx}$, $\hat{\Pi}_{xy}$, $\hat{\Delta}$ in this $\hat{u}_x$-$\hat{v}_x$-only problem. The coupled algorithm computes the non-conserve variables strictly under the complete mathematical constraint of the nonlinear constitutive model (NCCR). However, the uncoupled algorithm fails to recover the coupled solutions. This is because uncoupled algorithm splits the multi-dimensional problems into several non-interfering cases of one-dimensional problems. In the case of a three-dimensional problem, the stress and heat flux components $\left(\hat{\Pi}_{xx}, \hat{\Pi}_{xy}, \hat{\Pi}_{xz}, \hat{\Delta}, \hat{Q}_x\right)$ on a plane in the three-dimensional physical control volume induced by thermodynamic forces $\left(\hat{u}_x, \hat{v}_x, \hat{w}_x, \hat{T}_x\right)$ are computed directly by a linear summation of two one-dimensional solvers: (1) one on $\left(\hat{u}_x, 0, 0, \hat{T}_x\right)$, and (2) another on $\left(0, \hat{v}_x, 0, 0\right)$ or $\left(0, 0, \hat{w}_x, 0\right)$. In the $\hat{u}_x$-$\hat{v}_x$-only problem, the uncoupled solution process would neglect the interactional effect induced together by $\hat{u}_x$ and $\hat{v}_x$. Note that there is no more simple relation for the normal stresses $\hat{\Pi}_{xx}$, $\hat{\Pi}_{yy}$ and $\hat{\Pi}_{zz}$. The two solvers respectively for $\hat{u}_x$- and $\hat{v}_x$-induced flow in uncoupled algorithm cannot compute the stress $\hat{\Pi}_{yy}$ and $\hat{\Pi}_{zz}$. Therefore, an impression of the stress $\hat{\Pi}_{yy}$ and $\hat{\Pi}_{zz}$ only from NSF and coupled solver is shown in Figure 4. Furthermore, a two-dimensional and a three-dimensional compression-expansion problem ($\hat{u}_x$ - $\hat{v}_y$ -only and $\hat{u}_x$ - $\hat{v}_y$ - $\hat{w}_z$ -only) are also computed to highlight the uncoupled algorithm's deficiency in description of multi-dimensional effect. As is depicted in Figure 5, the uncoupled solutions in 1D, 2D, 3D compression-expansion problems are almost the same, while coupled algorithm succeeds in distinguishing multi-dimensional effect on the normal stress solutions.

Finally, we also compare inherent relation of computational consumption of these two algorithms for $\hat{u}_x$ -only, $\hat{v}_x$ -only and $\hat{u}_x$ - $\hat{v}_x$ -only problems with non-equilibrium level in Figure 6. The number of iteration in the vertical axis denotes the extra computational consumption in calculation of viscous flux term at every time-marching step and the horizontal axis represents the level of departure from equilibrium. Note that linear NSF stress and heat flux are computed explicitly from $\left(\hat{u}_x, \hat{v}_x, \hat{w}_x, \hat{T}_x\right)$. The computational consumption for NSF solver can be assumed to be unity as a criterion for NCCR solvers. As is depicted in Figure 6, the coupled algorithm's computation cost is lower than that of uncoupled algorithm in $\hat{u}_x$ -only and $\hat{u}_x$ - $\hat{v}_x$ -only problems. We can also see that the cost in the $\hat{v}_x$ -only problem is much lower than that of the uncoupled algorithm at local equilibrium regime, although it is more expensive at the away-from-equilibrium regime. On the whole, coupled algorithm improves the convergent speed in comparison with uncoupled algorithm.

## IV. Results and Discussion

The diatomic nonlinear constitutive model has gained some successful applications in shock wave structure [22, 24], micro-Couette flow [26], etc. In this paper, a significant objective is to extend this diatomic model into stable and efficient computation of high-speed flows past some three-dimensional complex configurations by utilizing our coupled solver. Meantime, we would also like to compare NSF and NCCR's capability of shockwave prediction in local non-equilibrium region in this section.

### A. Hypersonic flow past a blunted cone tip

The first case is a $25^0$ half angle cone with a blunted nose of 6.35 mm in radius. Since the wake region behind the cone is not of interest this time, the configuration is assumed to be infinitely long but be truncated to have its first 5 cm length from the leading edge as research object. Note that there is no angle attack for the free stream in this investigation and the cone tip is axisymmetric. An impression of 0.58 million structured grid used in all computations is given in Figure 7. The working gas is assumed to be pure nitrogen and the free-stream conditions come from the Run 31 of CUBRC experiments [27] with $U_\infty = 2764.5 m/s$ , $T_\infty = 144.4K$ , $p_\infty = 21.907 Pa$ , $Ma_\infty = 11.3$. An isothermal wall $T_w$=297.2$K$ and an accommodation coefficient of unity are assumed. The gas viscosity is computed by using the inverse power laws

$$\eta = \eta_{ref} \left( \frac{T}{T_{ref}} \right)^s . \tag{18}$$

And the other properties of nitrogen used in computations are given in Table 1

Figure 8-10 respectively demonstrate the DSMC, NSF and NCCR solutions of velocity and density along the lines normal to the cone at stagnation point, *X*= 1.14 and *X*= 3.14 cm. The DSMC data utilized in comparison come from Boyd's code (MONACO) and more details about the DSMC computation are introduced in [28]. The flow along the stagnation line passing through a normal shock wave is strongly nonequilibrium, where the linear NSF and DSMC solutions are therefore expected to be different to a large extent. As is depicted in Figure 8, the shock thickness and strength represented by NSF results are significantly different from the DSMC results in local region of cone tip. On the other hand, the NCCR results yield excellent agreement with the DSMC results and prove that the NCCR model improves the physical accuracy compared to NSF model.

Similar comparisons between DSMC and NCCR solutions for velocity and density at $X$= 1.14 and $X$= 3.14 cm are displayed in Figure 9-10. At these locations, the nonequilibrium effect from oblique shock wave is not as strong as that along the stagnation line, and the NSF model coupled with a slip boundary performs not too poorly. As shown in the figures, both NSF and NCCR results are very close to the DSMC results in most region, except NSF solutions in the oblique shock region. These discrepancies of NSF in velocity profiles are also highlighted in Figure 9-10(a). In general, the NCCR model performs better than the NSF model in predicting nonequilibrium flows.

**B. Complex flow past a hollow cylinder-flare**

Furthermore, before being employed to predict the complex flow of real vehicles in near-space regime which may be difficult for ground-based facilities, new computational models need to successfully meet some experimental verification firstly. In the testing case, the data of winds tunnel experiments made in the CUBRC LENS facility [29] is available to validate the computational model and this case has also been extensively studied in open literatures [30-33] containing a large amount of DSMC and NS validation data.

The configuration of a hollow cylinder in conjunction with a 30$^{o}$ conical flare is depicted in Figure 11. It is worthwhile mentioning that the sharp leading edge separates the free stream and only the external flow of this configuration is taken into account in this work as the internal flow does not interact with the external flow. The working diatomic gas is assumed pure nitrogen again. The specific inputs for the free-stream conditions are listed as

$$\begin{array}{ll} L = 0.1017m & U_\infty = 2301.7m/s \\ T_\infty = 118.2K & \rho_\infty = 9.023\times10^{-4}kg/m^3 \\ T_w = 295.6K & R = 296m^2/\left(\sec^2\cdot K\right) \\ Kn_\infty = 6.5\times10^{-4} & Ma_\infty = 10.4 \\ \eta_{ref} = 1.656\times10^{-5}\,\mathrm{N}\cdot s/m^2 & T_{ref} = 273K \end{array} . \quad (19)$$

Notice that the free-stream condition is non-equilibrium in CUBRC Run 14 as the translational, rotational and vibrational temperatures are at different time scales. However, since the NCCR model at present does not include these physical non-equilibrium effects, no transitional, rotational and vibrational energy exchange for a diatomic gas is considered in our simulation. According to Myong [15], the excess normal stress associated with the bulk viscosity of a diatomic gas could be introduced to describe simply the rotational non-equilibrium effect in some flow regimes where the rotational relaxation is faster than the hydrodynamic scale. Our present work here mainly focus on that in what range of flow regime NCCR model without these additional physical effect models can be capable of calculating.

Before we start to study the flow pass though the configuration, gradient-length-local Knudsen contour is shown firstly. As it is displayed in Figure 12, continuum breakdown occurs inside oblique shock wave above the flare and the region between sharp leading edge and the separation region. The size of the separation and re-attachment region can be observed clearly in Figure 13. The Mach number contour which is calculated by NCCR model demonstrates its capability of simulating these hypersonic flows fairly well.

Since the separation point is at about x/L=0.5, flow fields along the lines normal to the cylinder body at x/L=0.5 and 1 are studied. Figure 14 shows the detailed comparisons for density and velocity properties of DSMC, NSF and NCCR along these two lines. Notice that there is no comparison for temperature in this work now that the present NCCR model has not taken rotational and vibrational non-equilibrium effect into account. The DSMC and NSF results both come from the literature [31] and the DSMC results were also computed by a parallel optimized code named MONACO. As displayed on the left of Figure 14, the comparisons show that the NCCR results computed by the coupled iterative solver are in better agreement with the DSMC results than NSF results along the line normal to the cylinder at x/L=0.5 which is considered to be removed from local thermodynamic equilibrium based on continuum breakdown parameter $Kn_{GLL}$ . In the right profiles of Figure 14, NCCR results are only in qualitative agreement with DSMC results, but excellently capture a small discontinuity predicted by DSMC around the shock region which is not captured by NSF. Boyd, et al.[31] pointed out that DSMC did provide correct solution before the separation point, but the accuracy at the conjunction of the cylinder and flare(x/L=1) was questionable. The obvious deviation between NCCR and DSMC results makes it difficult to judge the performance of former one and its unknown cause remains worthwhile to be put on future's research. In Figure 15, numerical results are compared with the experimental data for non-dimensionalized heat flux coefficient along the body surface. The heat flux coefficient is defined as $C_q = q_w / 0.5\rho_\infty U_\infty^3$ . It is very encouraging that the NCCR solution is in outstanding agreement with the experimental data especially in the separation and re-attachment points (about x/L=0.6 and 1.4). In Figure 15, good performance of NCCR model in predicting the size of the separation and re-attachment region which is over-predicted by DSMC and under-predicted by NSF is also demonstrated. From the aforementioned analysis, the NCCR model is capable of simulating characteristic hypersonic flows fairly well.

### C. Slip flow past a HTV-type vehicle

In this case we would investigate the robust performance of our coupled solver when numerically simulating the hypersonic rarefied flows around 3D real vehicle's configuration in near-equilibrium or far-from-equilibrium flow regime. Flow fields of air with high speeds (Mach number 25 and 20) past a lifting body (a HTV-type vehicle), at the altitude of 50 and 90 kilometres are computed. Figure 16 shows the basic configuration of the HTV-type vehicle. Two attack angles ( $20^0$ and $0^0$ ) are considered in these two cases respectively. Both cases are assumed to have isothermal walls with a constant wall temperature 1000K. The viscosity of air is computed also by the inverse power laws (18) and the other properties of air used in computations are given in Table 2.

The 1st-order Maxwell-Smoluchowski slip boundary conditions with full tangential momentum and thermal accommodation coefficients are employed at the solid surface. The linear NSF with slip conditions is known to be capable of simulating these flows across the 2-meters-long HTV-type vehicle at altitudes of 50 and 90 kilometres, which could be defined as continuum and slip flow regimes respectively according to the body-length global (BLG) Knudsen number [34]

$$Kn_{BLG} = \frac{l_\infty}{L}, \tag{20}$$

where $l_\infty$ is the free-stream mean free path and $L$ is the characteristic body length. However, a non-negligible local rarefaction effect can be found especially in sharp leading edge, shock wave region or wake leeward region of hypersonic vehicles, which removes the flows away from local-equilibrium and out of the simulative capability of NSF model. Therefore, we adopt the second-order nonlinear model (NCCR) to simulate these flows as comparison, hoping it can make a significant supplement for linear NSF model with two aspects: recovering the NSF solution in continuum regime and remedying for the NSF solution in continuum breakdown regions.

In this work, grid independence study is carried on firstly using NCCR solution with three sets of grids. The details of these grids are present in Table 3. As illustrated in Figure 17, refined mesh 2 and 3 yield almost the same solution while coarse mesh 1 shows obvious deviation in Mach profile. Taking efficiency into consideration, we adopt mesh 2 for further investigation.

As a step toward testing the second-order nonlinear model's (NCCR's) potential in engineering application, the non-equilibrium regions of sharp leading edge, as well as surface aerothermal and aerodynamic properties including pressure, friction and heating transfer rate of the vehicle, are mainly investigated. The pressure, skin friction coefficient and heat transfer rate in Figure 23-24 are defined respectively as

$$C_p = \frac{p - p_\infty}{0.5\rho_\infty U_\infty^2}, \quad C_f = \frac{\tau}{0.5\rho_\infty U_\infty^2}, \quad C_h = \frac{q}{0.5\rho_\infty U_\infty^3}. \tag{21}$$

Figure 18 shows the local details of pressure contour and streamline pattern around the sharp leading edge on symmetrical plane, from which a strong shock wave structure and a weak re-attachment region can be observed clearly. In order to compare the locations of the shock wave structure around the edge predicted by NSF and NCCR model, we mainly focus on macro-variable distributions from three important positions ( $X = 0.1Rb$, $X = 0.5Rb$, $X = 1.0Rb$ ) along the y direction on the symmetrical plane. As illustrated in Figure 19, both NSF and NCCR can capture the steep shock structure accurately and its location is predicted exactly the same by these two models, which implies that NCCR model has as excellent performance as linear NSF model in continuum regime.

Figure 20 yields distribution profiles from another case at 90 kilometres high. It is worth mentioning that some significant discrepancies occur in these profiles between the near-local-equilibrium NSF results and the away-from-equilibrium NCCR results, e.g., the pressure values predicted by NSF are much higher than that by NCCR. In addition, Figure 21 shows the pressure fields predicted by the two models. Compared with that of Figure 18, the shock wave becomes weaker due to the lower density at 90 kilometres. Rarefaction effect becomes important and accounts for the discrepancies in these variable profiles.

Figure 22 gives an impression of the comparison of the after-body flows at different altitudes predicted by NS and NCCR. NCCR solution of after-body flow is consistent with NSF at 50 km, while the rarefied non-equilibrium effect starts to impact and causes obvious discrepancy between these two models' solutions at 90 km. Moreover, non-negligible discrepancies in the surface properties' profiles between NS and NCCR model are also revealed. A slight difference is shown in skin friction and heat transfer coefficient profiles in Figure 23. But the pressure, skin friction coefficients and heat transfer rate predicted by NCCR are much lower than those by NSF in Figure 24. NSF may overestimate the aerodynamic and aerothermal environment in rarefied flows with a strong non-equilibrium effect.

## V. Conclusions

The paper mainly focuses on the further extension of diatomic NCCR model into three-dimensional applications by using a robust and efficient coupled solver within finite volume framework. In our coupled algorithm which combines two conventional iterative methods' advantage, Fixed-point and Newton's iteration take turns to be employed during the computation process of NCCR model. The coupled algorithm not only overcomes uncoupled algorithm's shortcoming in description of multi-dimensional effect, but also improves the computational efficiency and stability. Based on the coupled solver for NCCR, numerical experiments have been carried out on three multi-dimensional cases, including pure nitrogen gas flow past a blunted cone tip, complex flow past the hollow cylinder-flare and slip flows past a HTV-type lifting-body configuration at 50 and 90 kilometres. Overall, the NCCR model can recover to NSF solution in continuum regime and yields better agreement with the DSMC and experiment data than NSF in local non-equilibrium regions. We can also see that the discrepancies of flow field and surface properties between NSF and NCCR solutions are enlarged as the rarefied nonequilibrium effect is strengthened. On the whole, it can be concluded from the certification and validation with DSMC and experiment, that NCCR model has the potential as an alternative to less accurate NSF linear constitutive relations in the prediction of hypersonic and rarefied flows.

## VI. Acknowledgements

The first author of this paper (Zhongzheng Jiang) was supported by the China Scholarship Council (grant number 201706320214). This research was fund by the National Natural Science Foundation of China (Grant NO.11502232, NO.51575487 and NO.11572284), the National Basic Research Program of China (Grant NO. 2014CB340201) and the Fundamental Research Funds for the Central Universities.

## VII. References

1. Bird, G. A. *Molecular gas dynamics and the direct simulation of gas flows*. Oxford: Clarendom Press, 1994.
2. Grad, H. "Asymptotic theory of the Boltzmann equation," *Physics of Fluids* Vol. 6, No. 2, 1963, pp. 147-181.doi: 10.1063/1.1706716
3. Bhatnagar, P. L., Gross, E. P., and Krook, M. "A model for collision processes in gases. I. Small amplitude processes in charged and neutral one-component systems," *Physical Review* Vol. 94, No. 3, 1954, pp. 511-525.doi: 10.1103/PhysRev.94.511
4. Holway, L. H. "New statistical models for kinetic theory: methods of construction," *The Physics of Fluids* Vol. 9, No. 9, 1966, pp. 1658-1673.doi: 10.1063/1.1761920
5. Shakhov, E. M. "Generalization of the Krook kinetic relaxation equation," *Fluid Dynamics* Vol. 3, No. 5, 1968, pp. 95-96.doi: 10.1007/BF01029546
6. Broadwell, J. E. "Study of rarefied shear flow by the discrete velocity method," *Journal of Fluid Mechanics* Vol. 19, No. 03, 2006, p. 401.doi: 10.1017/s0022112064000817
7. Xu, K., and Huang, J.-C. "A unified gas-kinetic scheme for continuum and rarefied flows," *Journal of Computational Physics* Vol. 229, No. 20, 2010, pp. 7747-7764.doi: 10.1016/j.jcp.2010.06.032
8. Burnett, D. "The Distribution of Molecular Velocities and the Mean Motion in a Non-Uniform Gas," *Proceedings of the London Mathematical Society* Vol. s2-40, No. 1, 1936, pp. 382-435. doi: 10.1112/plms/s2-40.1.382
9. Zhong, X., Maccormack, R. W., and Chapman, D. R. "Stabilization of the Burnett equations and application to hypersonicflows," *AIAA Journal* Vol. 31, No. 6, 1993, pp. 1036-1043. doi: 10.2514/3.11726
10. Zhao, W., Chen, W., and Agarwal, R. K. "Formulation of a new set of Simplified Conventional Burnett equations for computation of rarefied hypersonic flows," *Aerospace Science and Technology* Vol. 38, 2014, pp. 64-75.doi: http://dx.doi.org/10.1016/j.ast.2014.07.014
11. Grad, H. "On the kinetic theory of rarefied gases," *Comm.Pure Appl. Math* Vol. 2, 1949, pp. 331-407.
12. Torrilhon, M. "Modeling nonequilibrium gas flow based on moment equations," *Annual Review of Fluid Mechanics* Vol. 48, No. 1, 2016, pp. 429-458.doi: 10.1146/annurev-fluid-122414-034259
13. Eu, B. C. *Kinetic theory and irreversible thermodynamics*. New York: Wiley, 1992.
14. Myong, R. S. "Thermodynamically consistent hydrodynamics computational models for high-Knudsen-number gas flows," *Physics of Fluids* Vol. 11, No. 9, 1999, pp. 2788-2802. doi: 10.1063/1.870137
15. Myong, R. S. "A generalized hydrodynamic computational model for rarefied and microscale diatomic gas flows," *Journal of Computational Physics* Vol. 195, No. 2, 2004, pp. 655-676. doi: 10.1016/j.jcp.2003.10.015
16. Rana, A., Ravichandran, R., Park, J. H., and Myong, R. S. "Microscopic molecular dynamics characterization of the second-order non-Navier-Fourier constitutive laws in the Poiseuille gas flow,"

*Physics of Fluids* Vol. 28, No. 8, 2016, p. 082003. doi: 10.1063/1.4959202
17. Myong, R. S. "A computational method for Eu's Generalized Hydrodynamic Equations of rarefied and microscale gasdynamics," *Journal of Computational Physics* Vol. 168, No. 1, 2001, pp. 47-72. doi: 10.1006/jcph.2000.6678
18. Jiang, Z., Zhao, W., and Chen, W. "A three-dimensional finite volume method for conservation laws in conjunction with modified solution for nonlinear coupled constitutive relations," *the 30th International Symposium on Rarefied Gas Dynamics*. Vol. 1786, American Institute of Physics, Victoria, British Columbia, Canada, 2016, p. 040002.
19. Jiang, Z., Chen, W., and Zhao, W. "Numerical simulation of three-dimensional high-speed flows using a second-order nonlinear model." 2017, p. 2346.
20. Eu, B. C. *Generalized thermodynamics: The thermodynamics of irreversible processes and generalized hydrodynamics*. New York: Kluwer Academic Publishers, 2002.
21. Mazen Al-Ghoul, and Eu, B. C. "Generalized hydrodynamics and shock waves," *Physical Review E* Vol. 56, 1997, pp. 2981-2992.doi: 10.1103/PhysRevE.56.2981
22. Eu, B. C., and Ohr, Y. G. "Generalized hydrodynamics, bulk viscosity, and sound wave absorption and dispersion in dilute rigid molecular gases," *Physics of Fluids* Vol. 13, No. 3, 2001, pp. 744-753. doi: 10.1063/1.1343908
23. Myong, R. S. "On the high Mach number shock structure singularity caused by overreach of Maxwellian molecules," *Physics of Fluids* Vol. 26, No. 5, 2014, p. 056102.
24. Zhao, W., Jiang, Z., and Chen, W. "Computation of 1-D shock structure using nonlinear coupled constitutive relations and generalized hydrodynamic equations," *the 30th International Symposium on Rarefied Gas Dynamics*. Vol. 1786, American Institute of Physics, Victoria, British Columbia, Canada, 2016, p. 140007.
25. Kim, K. H., Kim, C., and Rho, O.-H. "Methods for the accurate computations of hypersonic flows: I. AUSMPW+ scheme," *Journal of Computational Physics* Vol. 174, No. 1, 2001, pp. 38-80.
26. Jiang, Z., Chen, W., and Zhao, W. "Numerical analysis of the micro-Couette flow using a non-Newton–Fourier model with enhanced wall boundary conditions," *Microfluidics and Nanofluidics* Vol. 22, No. 1, 2017, p. 10. doi: 10.1007/s10404-017-2028-y
27. Holden, M. "Experimental database from CUBRC studies in hypersonic laminar and turbulent interacting flows including flowfield chemistry," *Prepared for RTO code validation of DSMC and Navier-Stokes code validation studies, Calspan-University at Buffalo Research Center, Buffalo, NY*, 2000, pp. 105-114.
28. Wang, W.-L., and Boyd, I. "Hybrid DSMC-CFD simulations of hypersonic flow over sharp and blunted bodies," 2003.
29. Holden, M. S. "Measurement in regions of laminar shock wave/boundary layer interaction in hypersonic flow-code validation." CUBRC Report, 2003.
30. Harvey, J. K. "A Review of a Validation Exercise on the use of the DSMC Method to Compute Viscous/Inviscid Interactions in Hypersonic Flow," *36th AIAA Thermophysics Conference*. Orlando, Florida, 2003.
31. Wang, W.-L., and Boyd, I. "Hybrid DSMC-CFD Simulations of Hypersonic Flow over Sharp and Blunted Bodies," *36th AIAA Thermophysics Conference*. Orlando, Florida, 2003, pp. 1-13.
32. Jiang, T., Xia, C., and Chen, W. "An improved hybrid particle scheme for hypersonic rarefied-continuum flow," *Vacuum* Vol. 124, 2016, pp. 76-84. doi: 10.1016/j.vacuum.2015.11.012
33. Zhao, W. "Linearized stability analysis and numerical computation of Burnett equations in hypersonic

flow." Zhejiang University, 2014.

34. Iain D.Boyd, G. C., Graham V. Candler. "Predicting failure of the continuum fluid equations in transitional hypersonic flows," *Physics of Fluids* Vol. 7, No. 1, 1995, pp. 210-219. doi: 10.1063/1.868720

Table 1 Physical gas properties of pure nitrogen gas

| $\gamma$ | $Pr$ | $R(\mathrm{J/kg\cdot K})$ | $c$ | $f_b$ | $T_{ref}\,(\mathrm{K})$ | $\eta_{ref}\,(\mathrm{Pa\cdot s})$ | $s_{VHS}$ |
|---|---|---|---|---|---|---|---|
| 1.4 | 0.72 | 296.7 | 1.02029 | 0.8 | 273 | $1.656\times10^{-5}$ | 0.74 |

Table 2: Physical gas properties of air

| $\gamma$ | $Pr$ | $R(\mathrm{J/kg\cdot K})$ | $c$ | $f_b$ | $T_{ref}\,(\mathrm{K})$ | $\eta_{ref}\,(\mathrm{Pa\cdot s})$ | $s_{VHS}$ |
|---|---|---|---|---|---|---|---|
| 1.4 | 0.72 | 287.1 | 1.01445 | 0.8 | 273.15 | $1.71608\times10^{-5}$ | 0.77 |

Table 3 Grid independence study

| | Grid points in normal, flow and circumference directions | Grid spacing nearest to the solid wall |
|---|---|---|
| Mesh 1 | $60\times120\times140$ | $5\times10^{-5}$ |
| Mesh 2 | $130\times150\times160$ | $5\times10^{-5}$ |
| Mesh 3 | $180\times170\times200$ | $1\times10^{-5}$ |

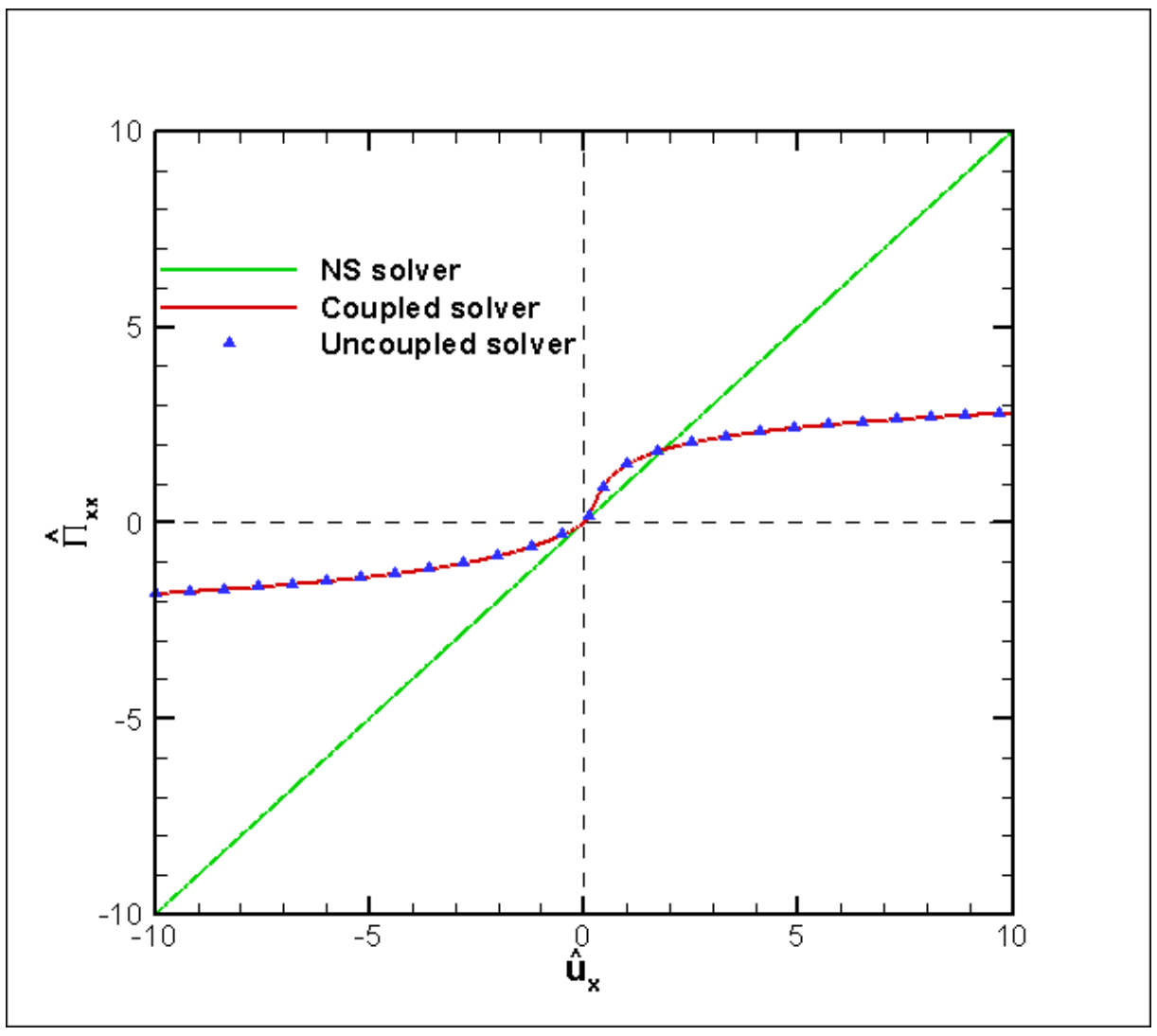

(a) $\hat{\Pi}_{xx}$ varies with $\hat{u}_x$

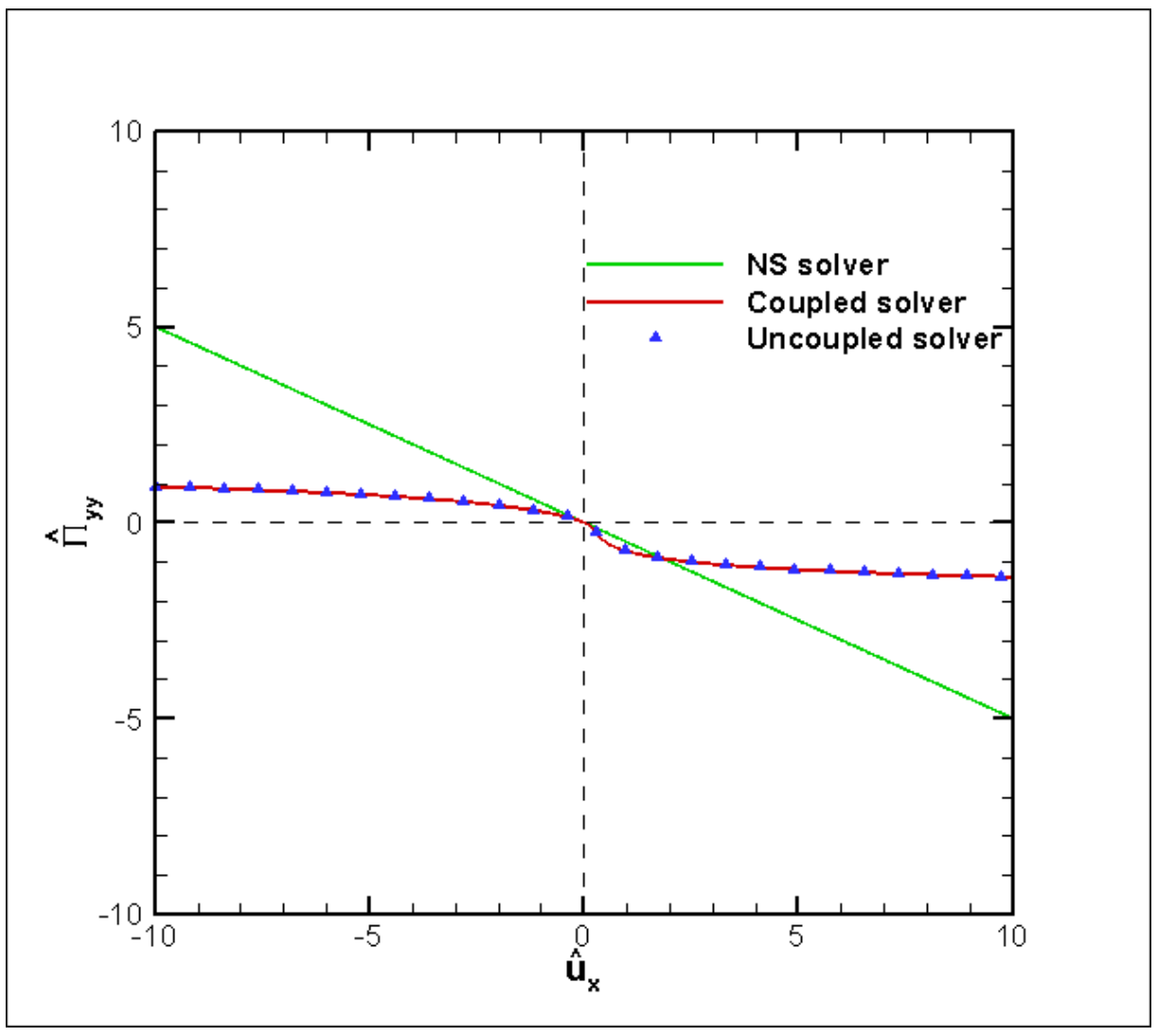

(b) $\hat{\Pi}_{yy}$ varies with $\hat{u}_x$

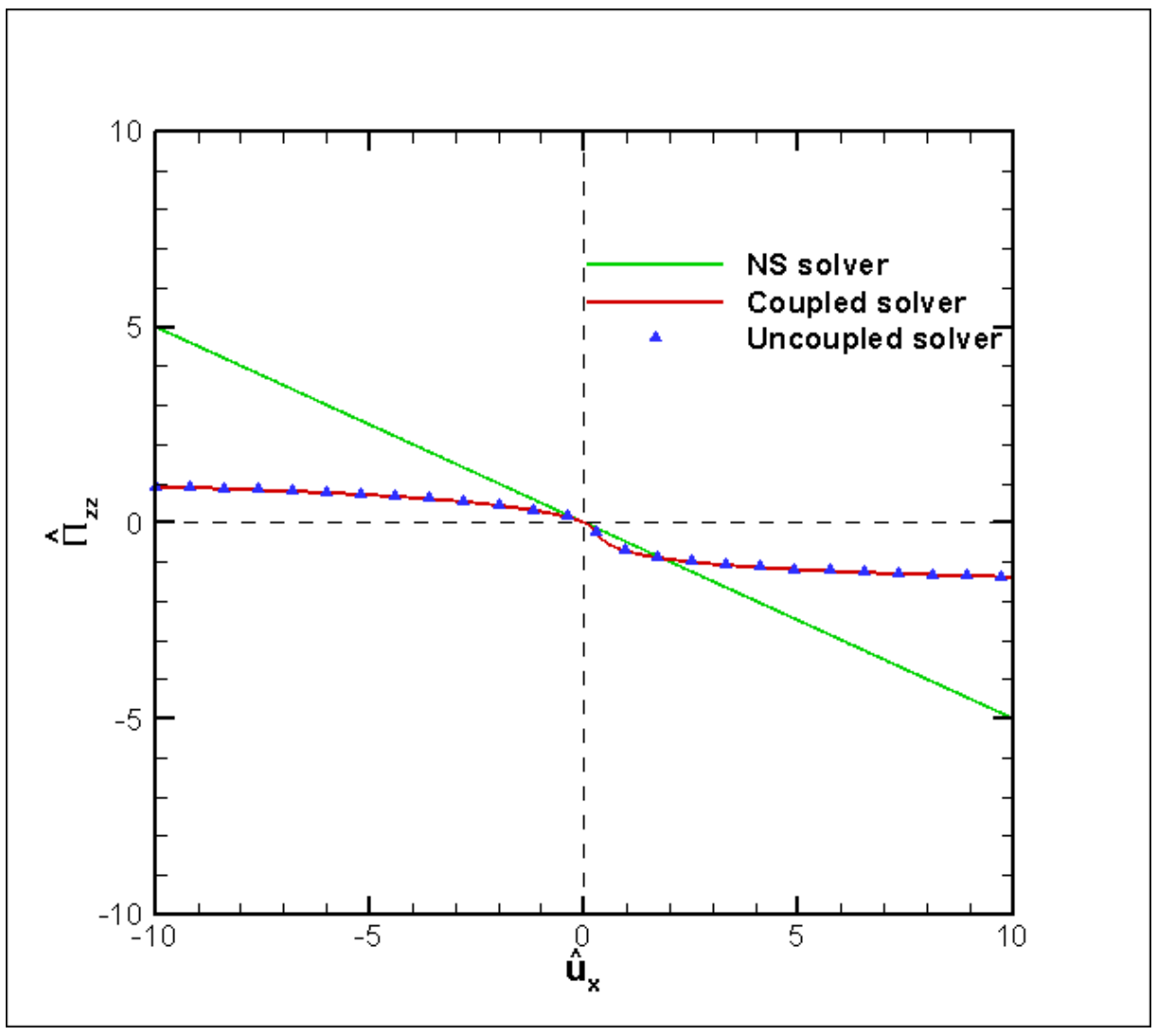

(c) $\hat{\Pi}_{zz}$ varies with $\hat{u}_x$

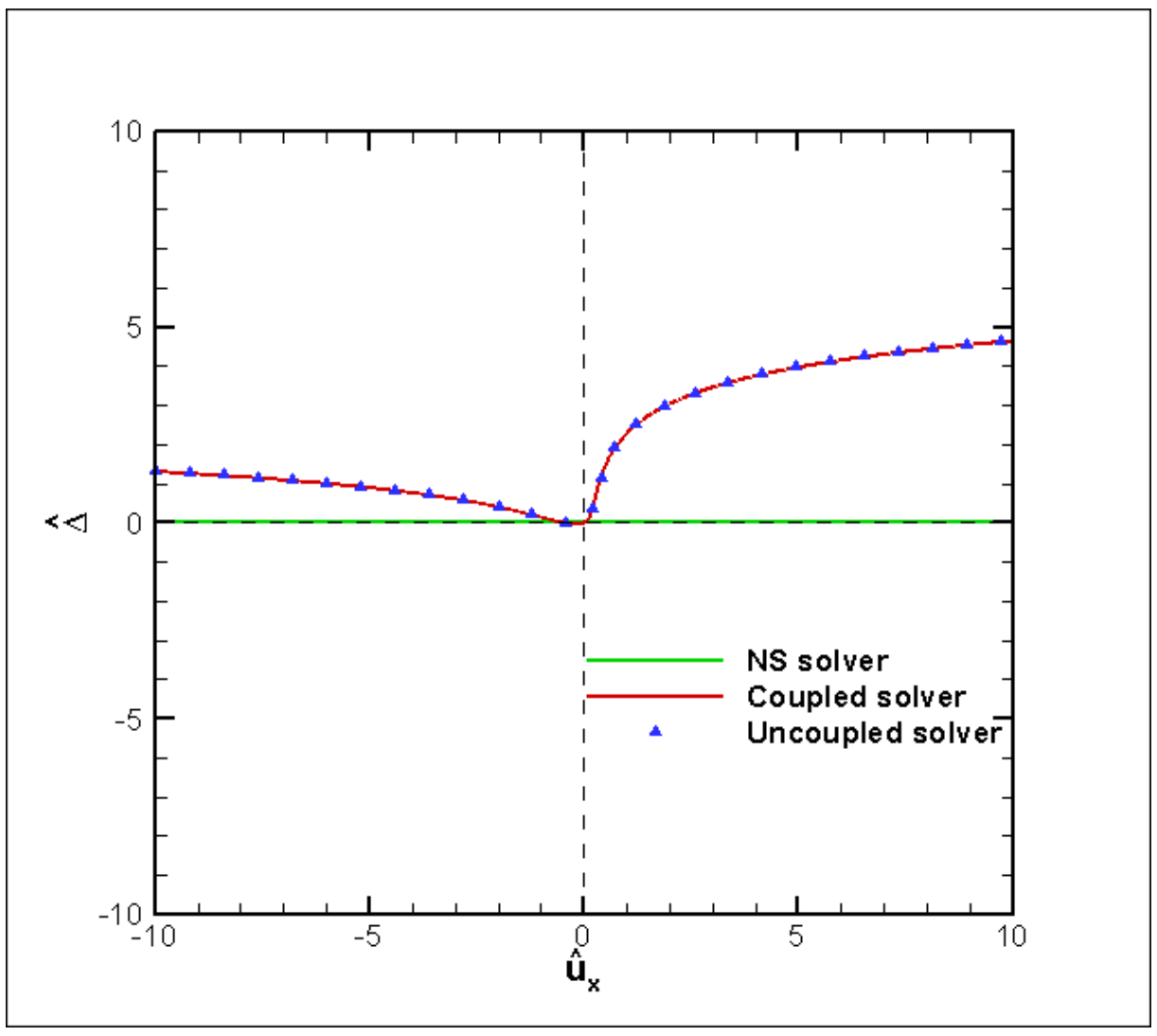

(d) $\hat{\Delta}$ varies with $\hat{u}_x$

Figure 1 Constitutive relations in the $\hat{u}_x$ -only problem for a diatomic gas

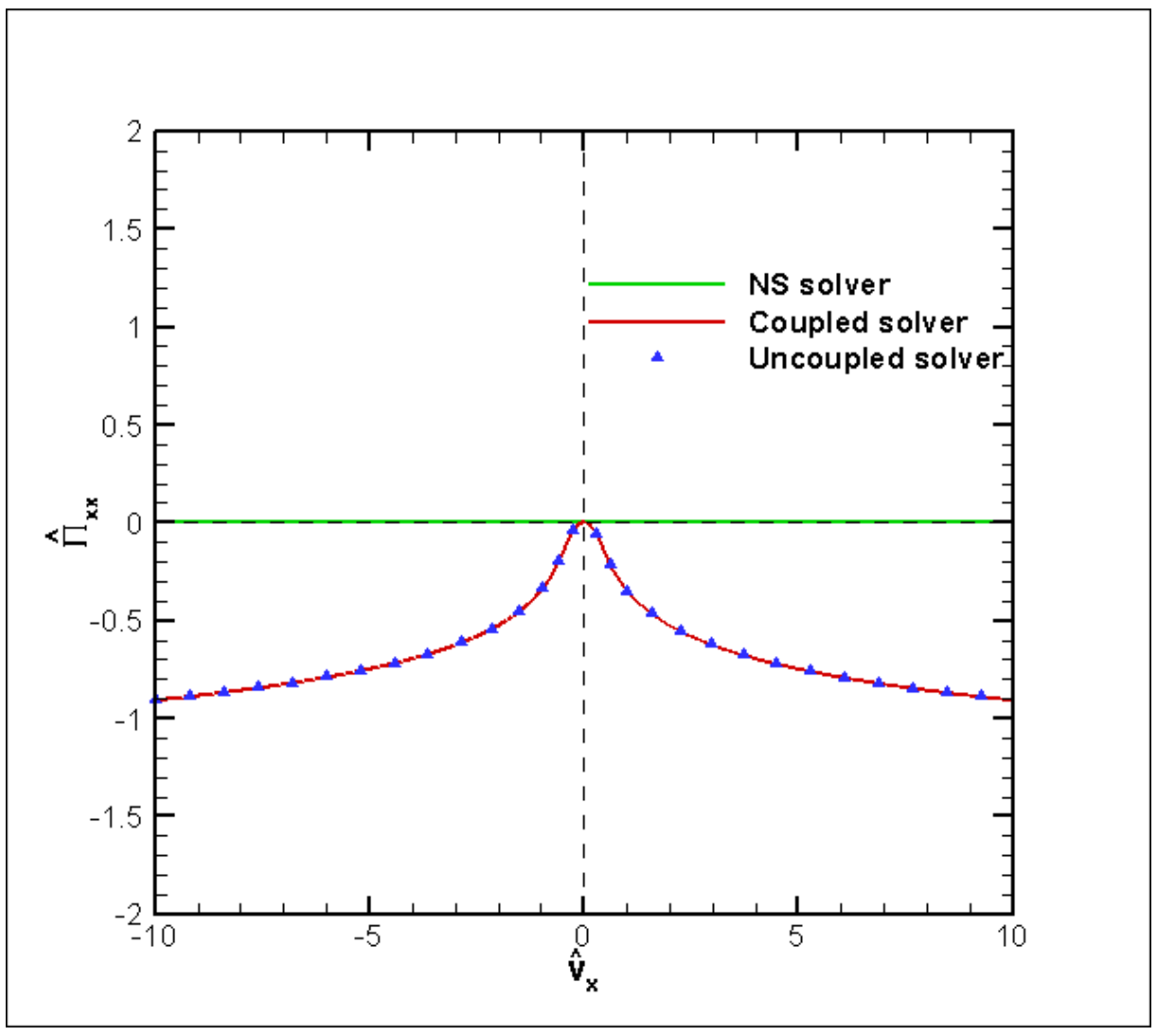

(a) $\hat{\Pi}_{xx}$ varies with $\hat{v}_x$

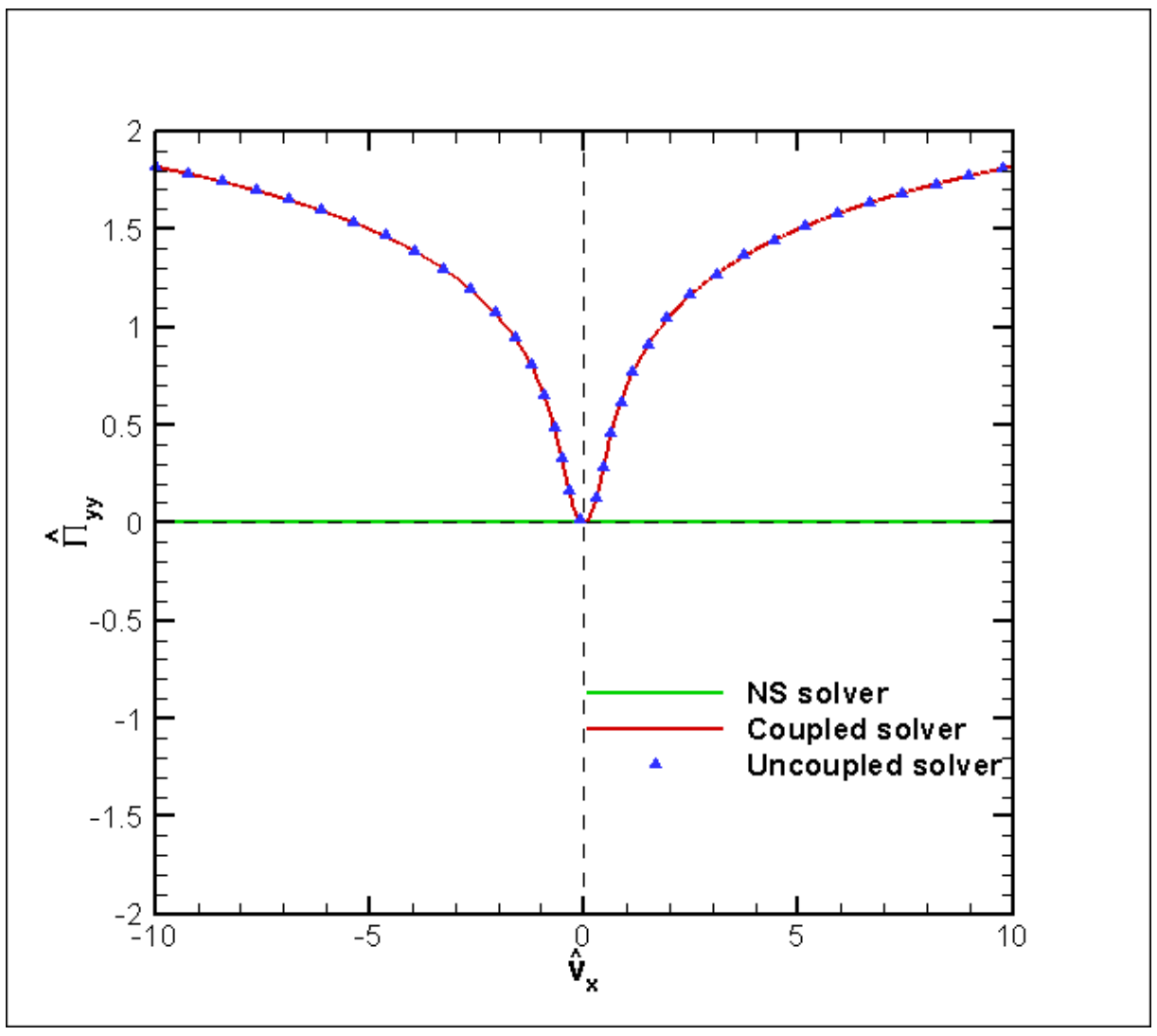

(b) $\hat{\Pi}_{yy}$ varies with $\hat{v}_x$

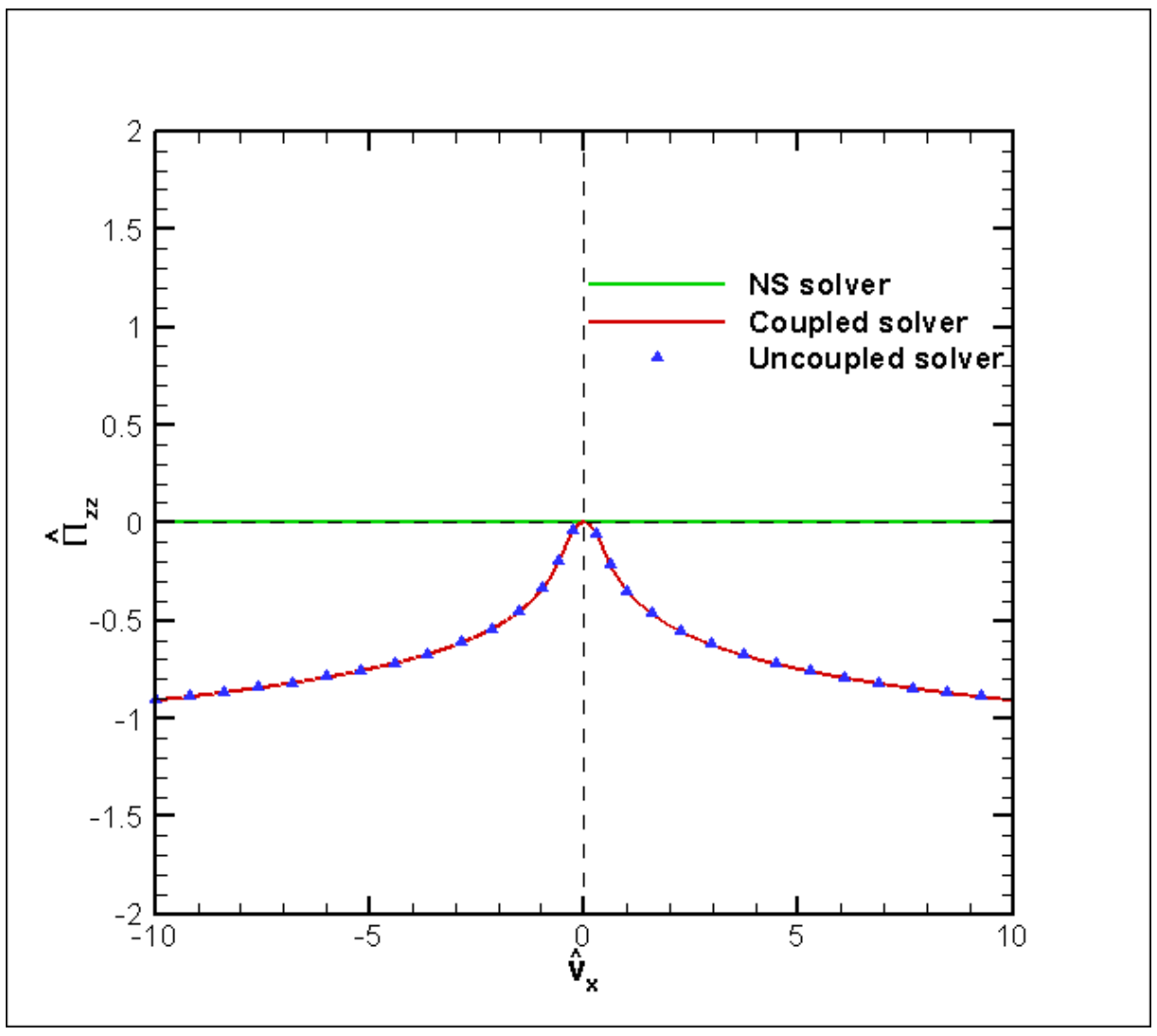

(c) $\hat{\Pi}_{zz}$ varies with $\hat{v}_x$

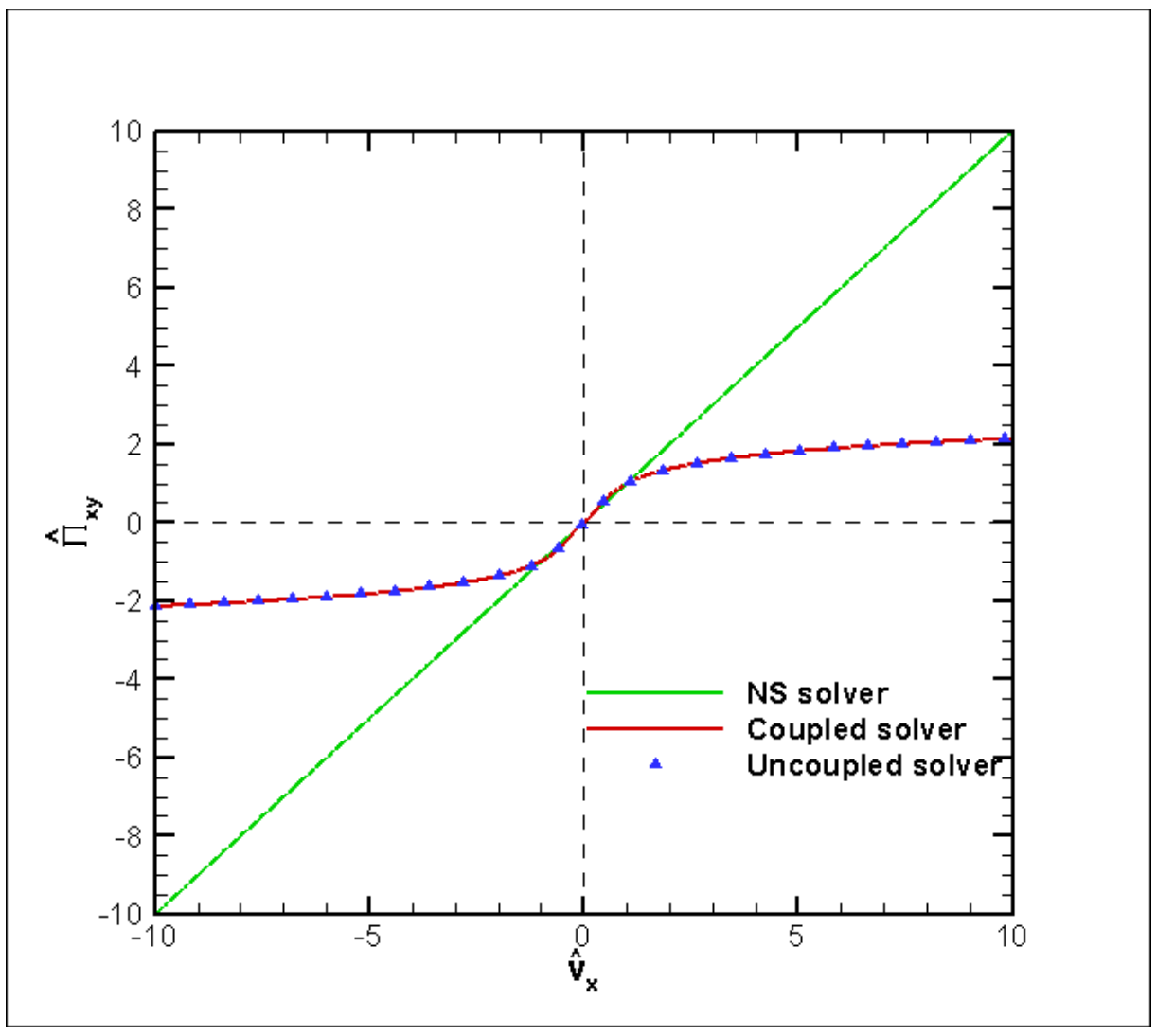

(d) $\hat{\Pi}_{xy}$ varies with $\hat{v}_x$

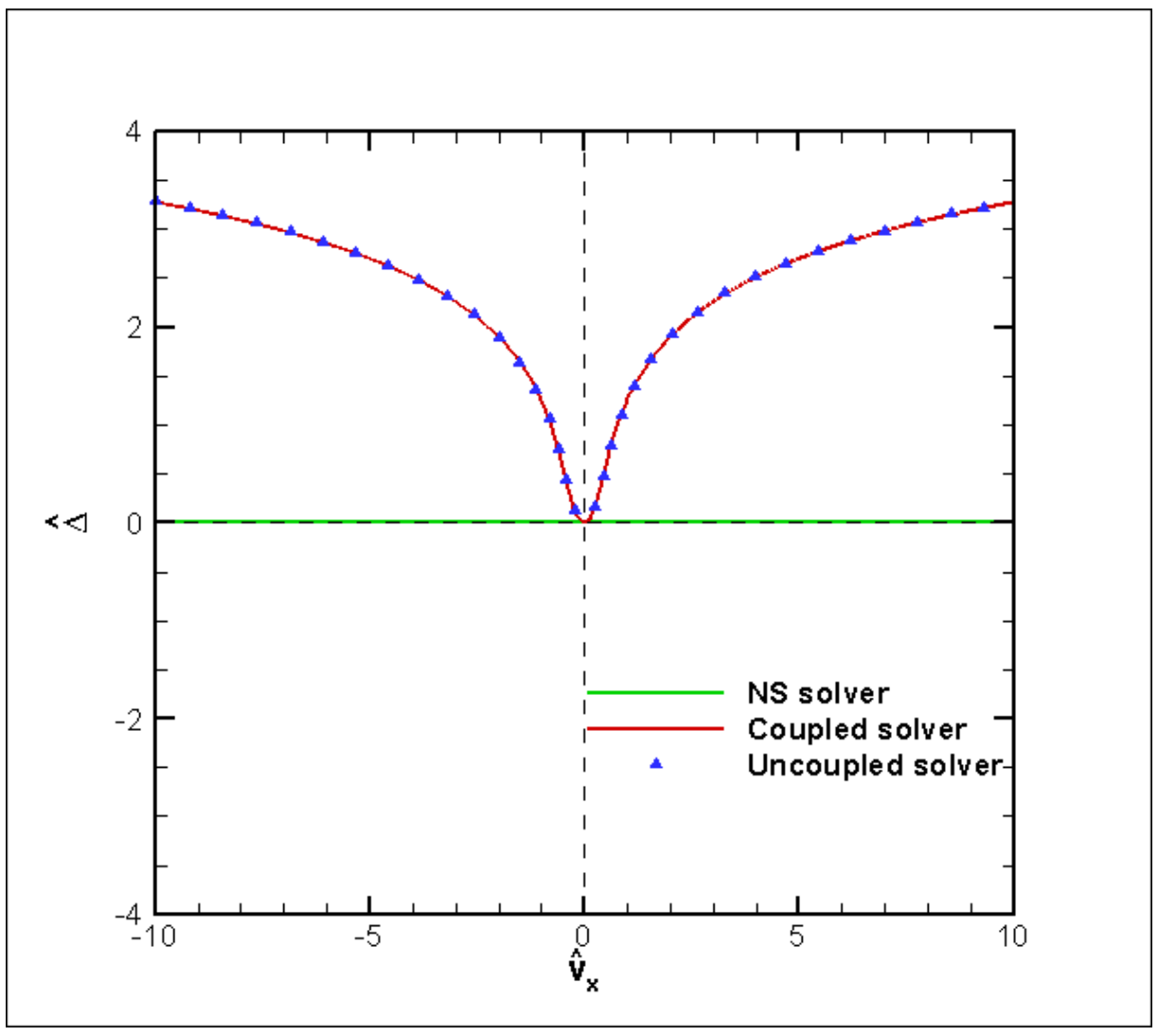

(e) $\hat{\Pi}_{xx}$ varies with $\hat{v}_x$

Figure 2 Constitutive relations in the $\hat{v}_x$ -only problem for a diatomic gas

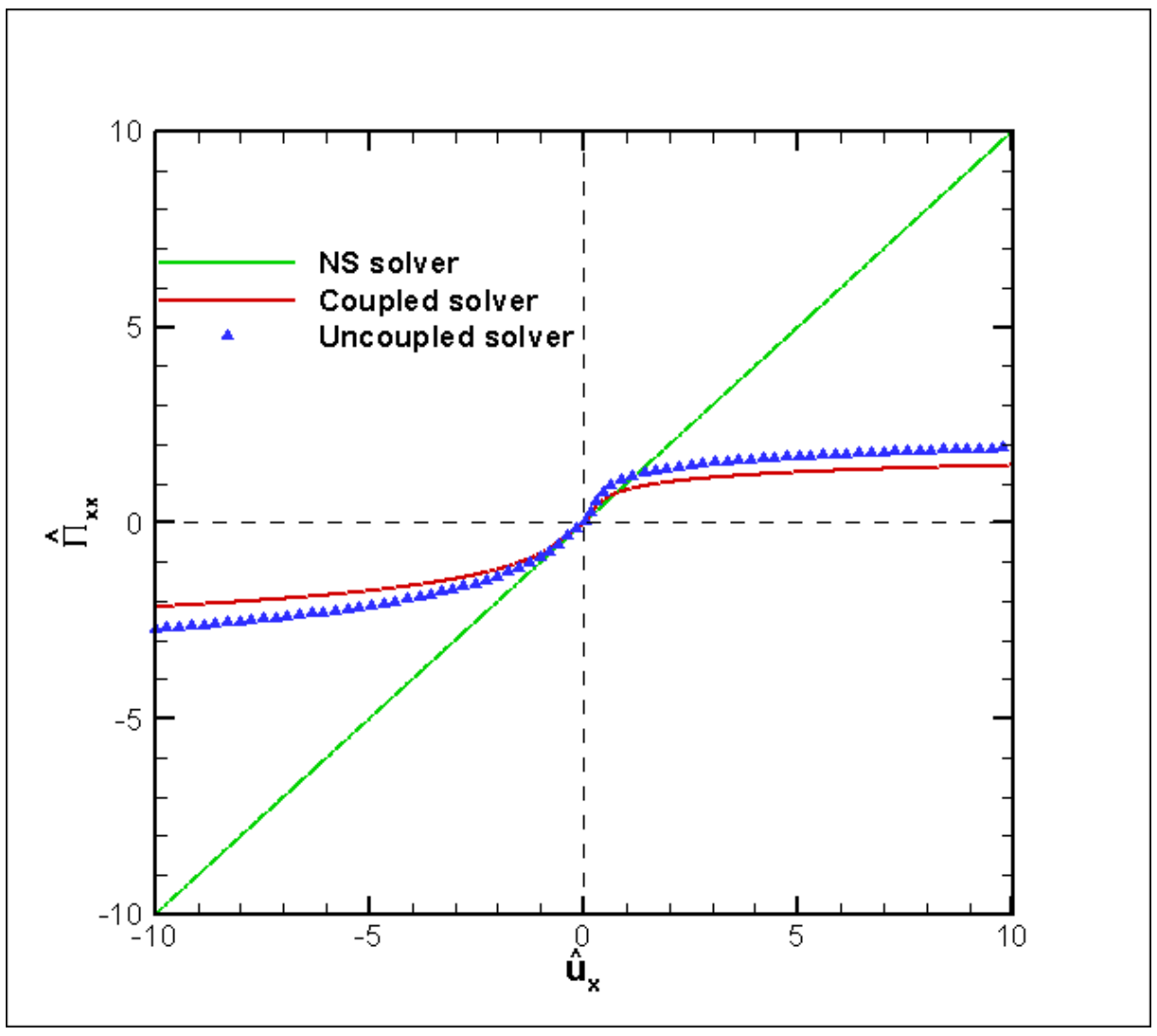

(a) $\hat{\Pi}_{xx}$ varies with $\hat{u}_x$

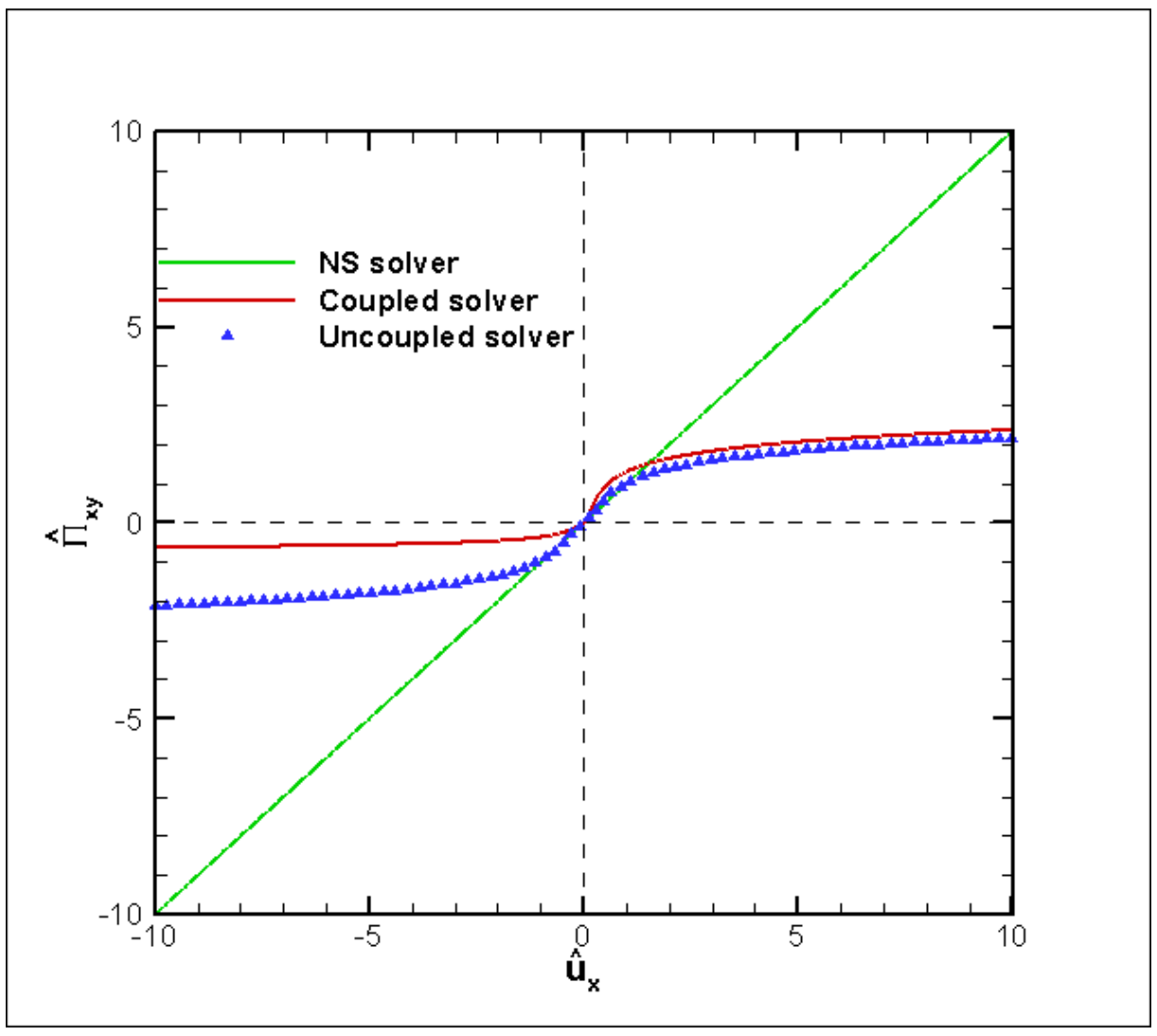

(b) $\hat{\Pi}_{xy}$ varies with $\hat{u}_x$

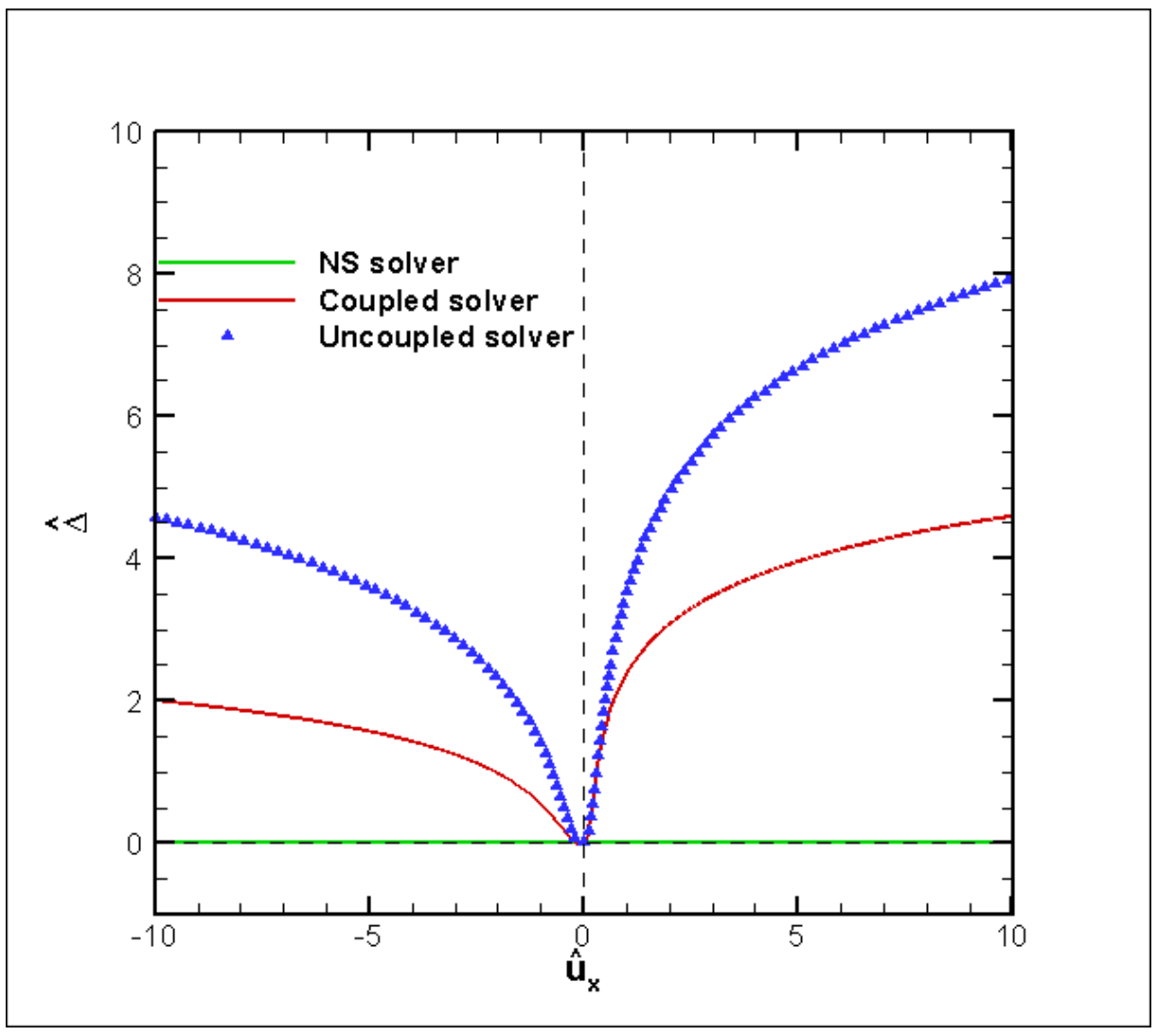

(c) $\hat{\Delta}$ varies with $\hat{u}_x$

Figure 3 Nonlinear constitutive relations relative to the Navier-Stokes relations in the $\hat{u}_x$ - $\hat{v}_x$ -only problem for a diatomic gas

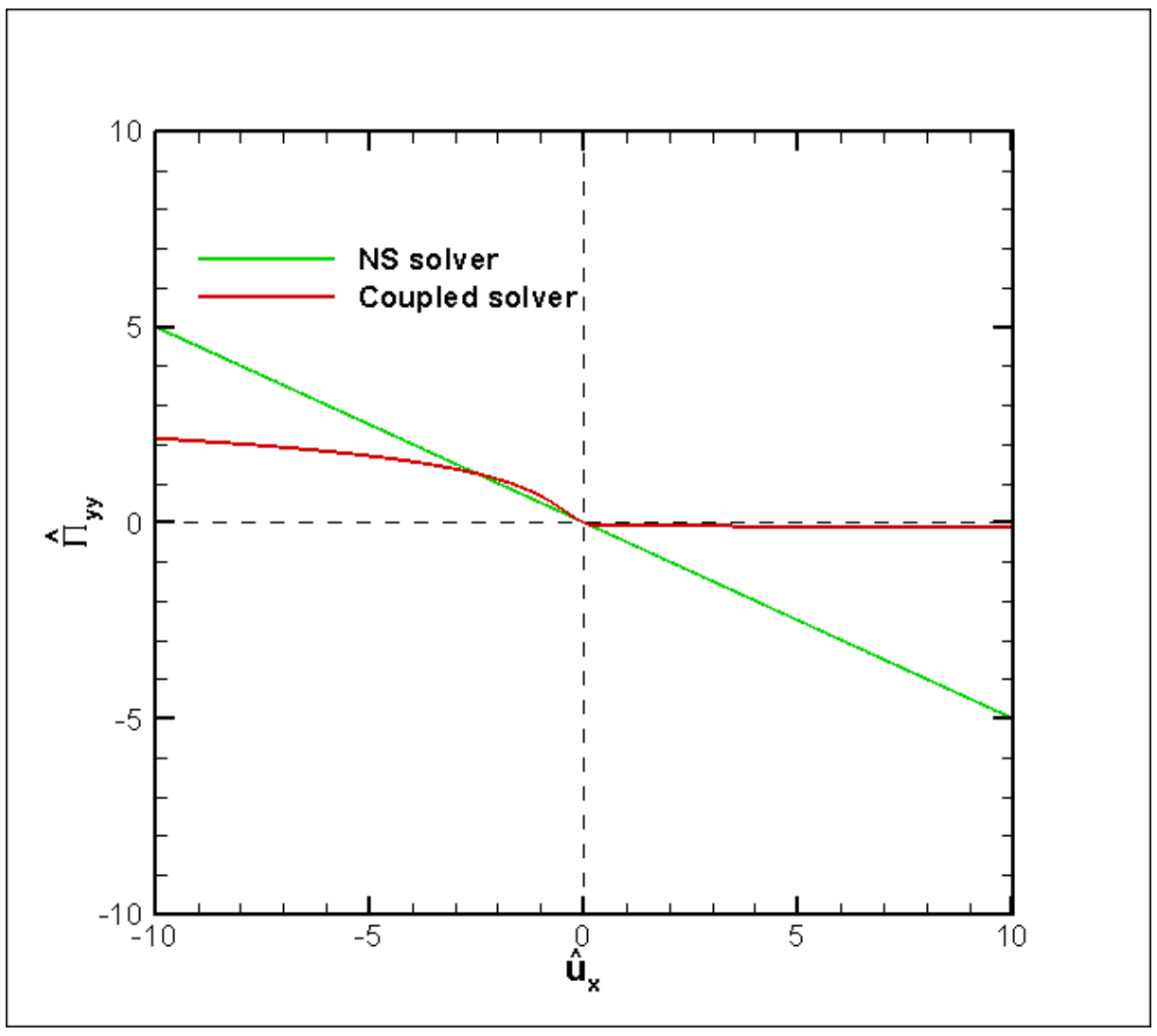

(a) $\hat{\Pi}_{yy}$ varies with $\hat{u}_x$

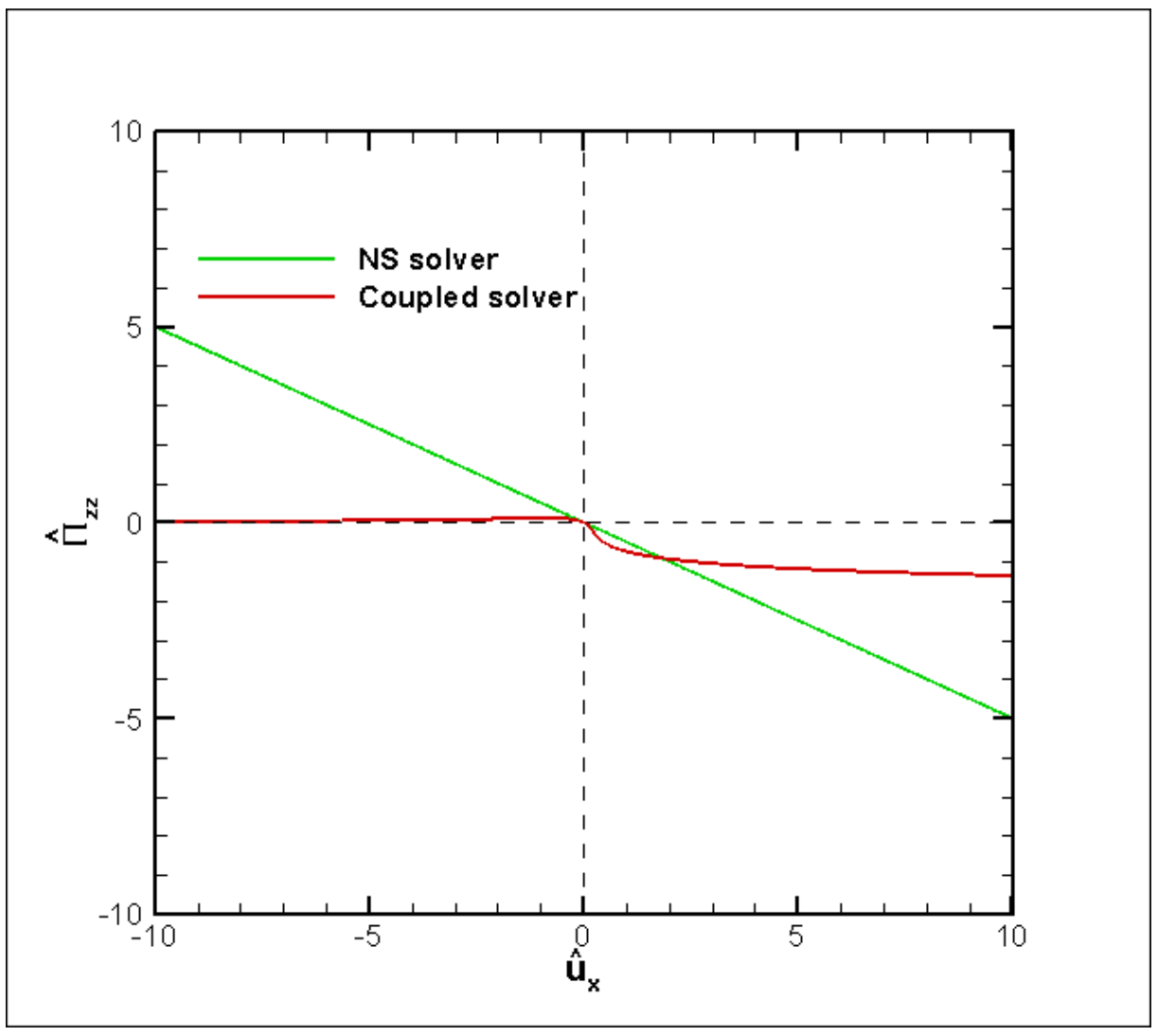

(b) $\hat{\Pi}_{zz}$ varies with $\hat{u}_x$

Figure 4 Nonlinear constitutive relations $\hat{\Pi}_{yy}$ (left) and $\hat{\Pi}_{zz}$ (right) with $\hat{u}_x$ computed by coupled solver relative to the Navier-Stokes relations in the $\hat{u}_x$ - $\hat{v}_x$ -only problem for a diatomic gas

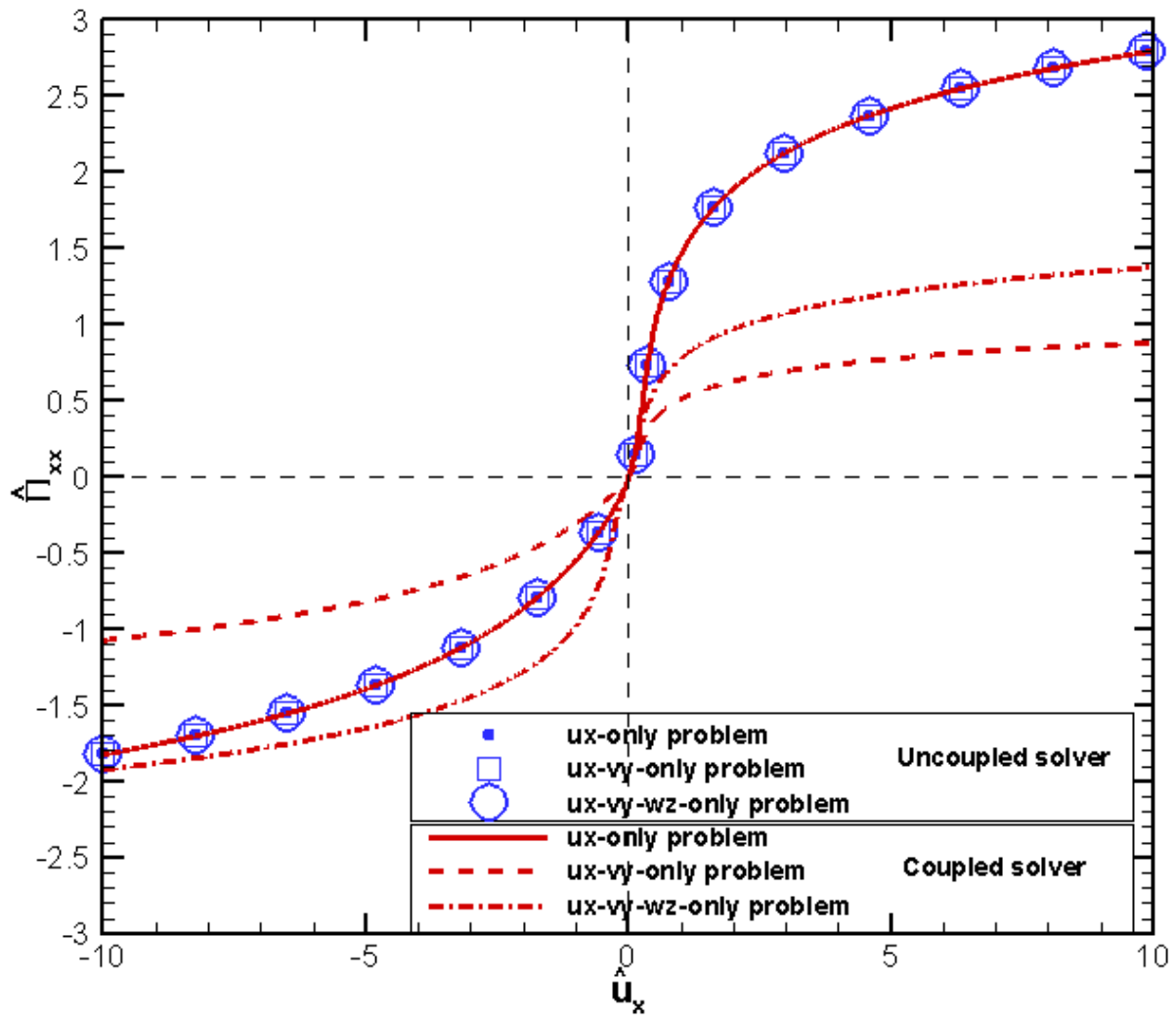

Figure 5 The nonlinear relations between $\hat{\Pi}_{xx}$ and $\hat{u}_x$ computed by uncoupled and coupled solver in the multi-dimensional compression and expansion problem for a diatomic gas

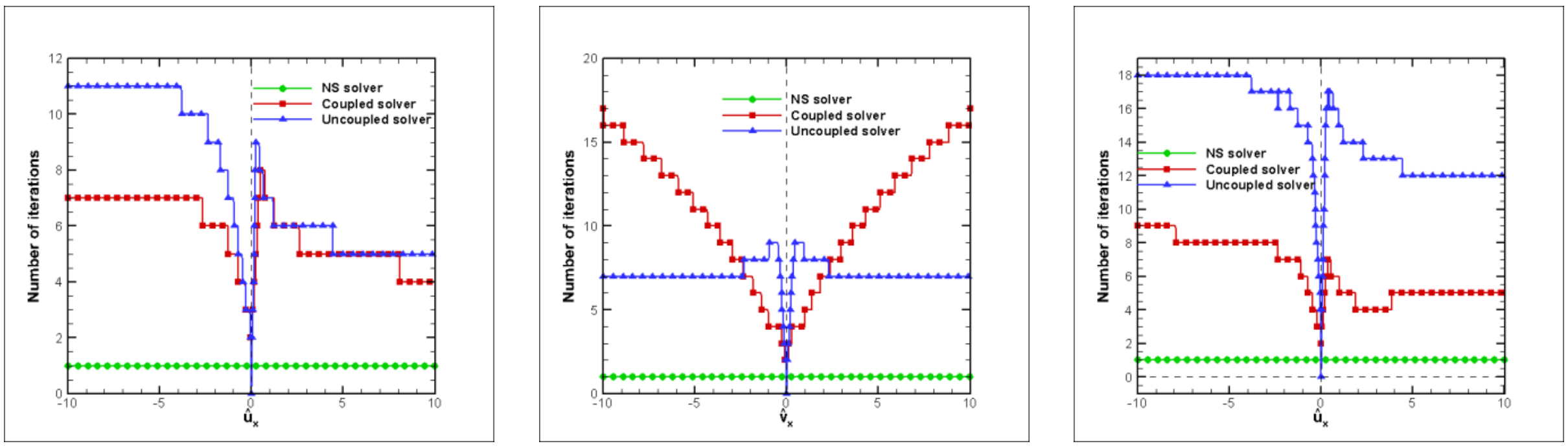

(a) $\hat{u}_x$ -only problem (b) $\hat{v}_x$ -only problem (c) $\hat{u}_x$ - $\hat{v}_x$ -only problem

Figure 6 Number of iterations for the calculation of non-conservative variables for a diatomic gas

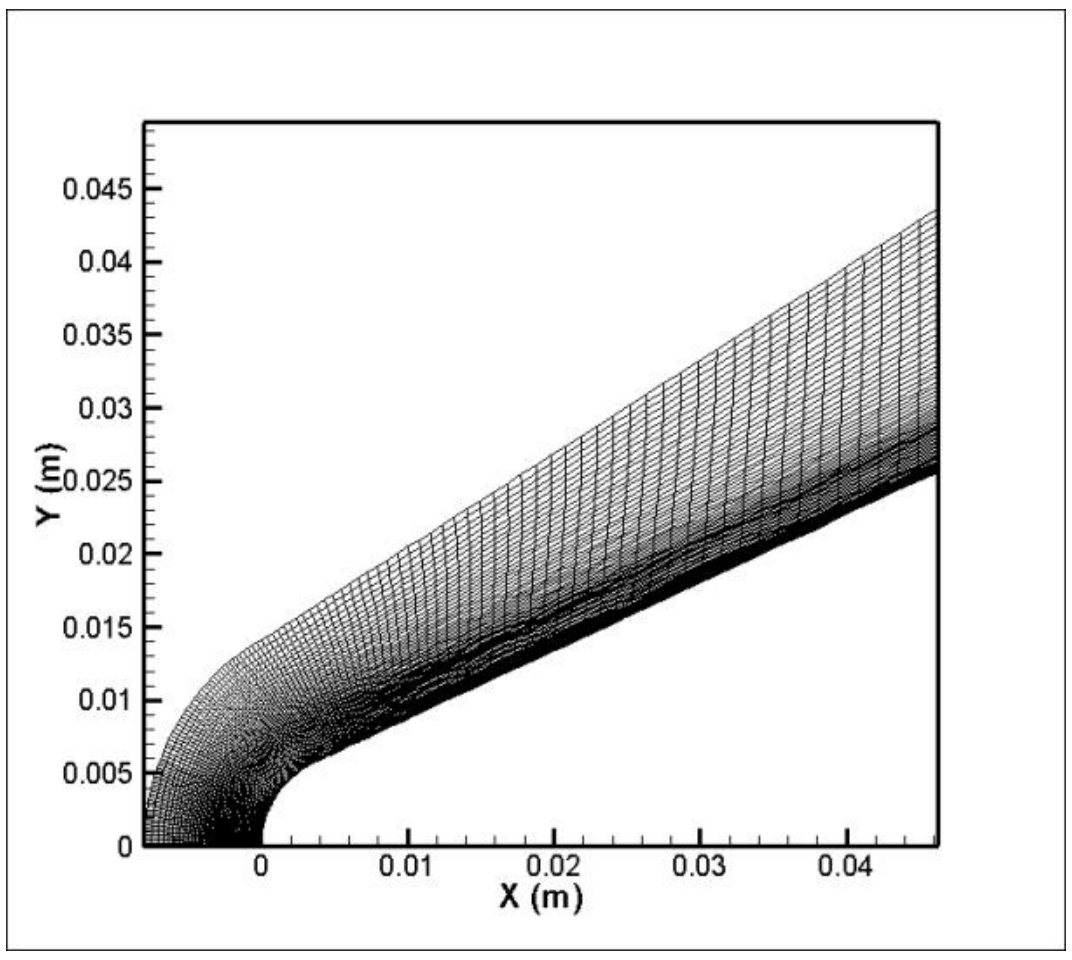

Figure 7 Structured grid of the $25^0$ blunted cone's symmetrical plane.

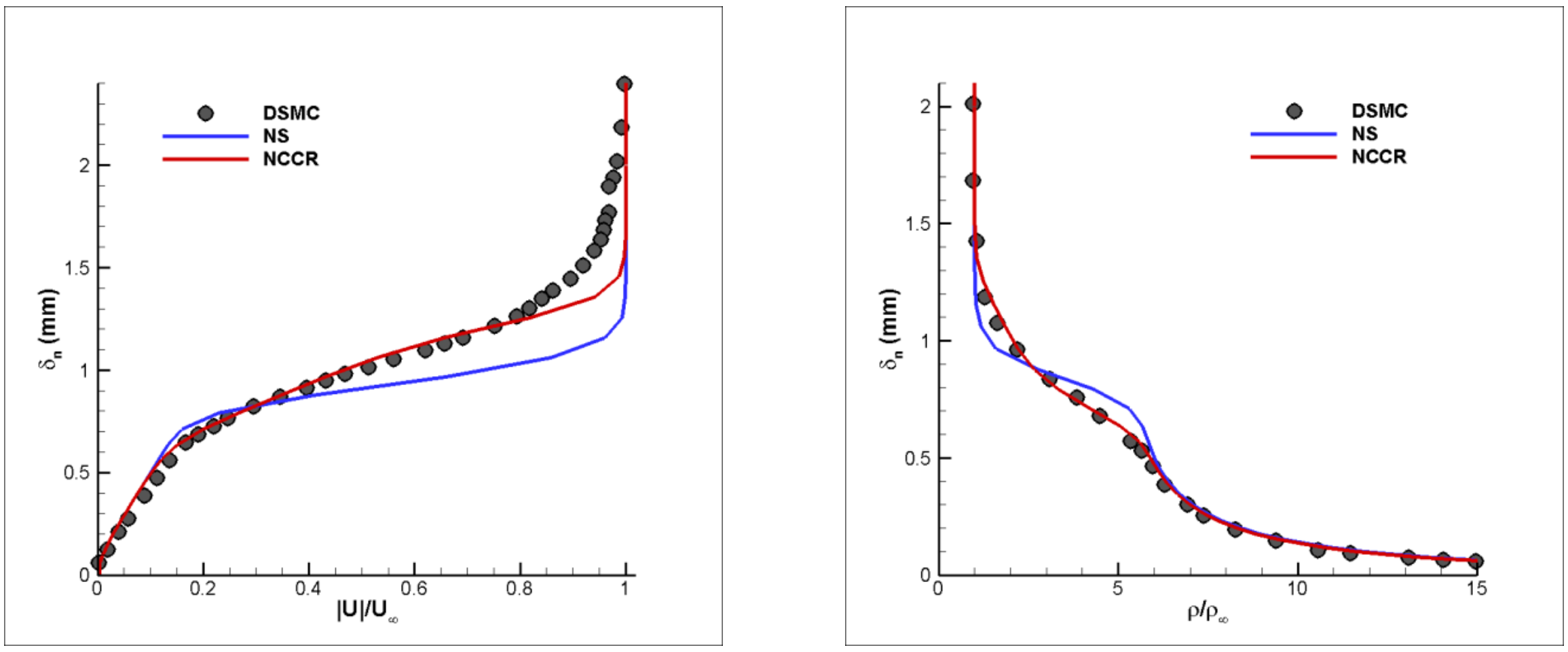

(a) Velocity (b) Density

Figure 8 Profiles along the line normal to the stagnation point.

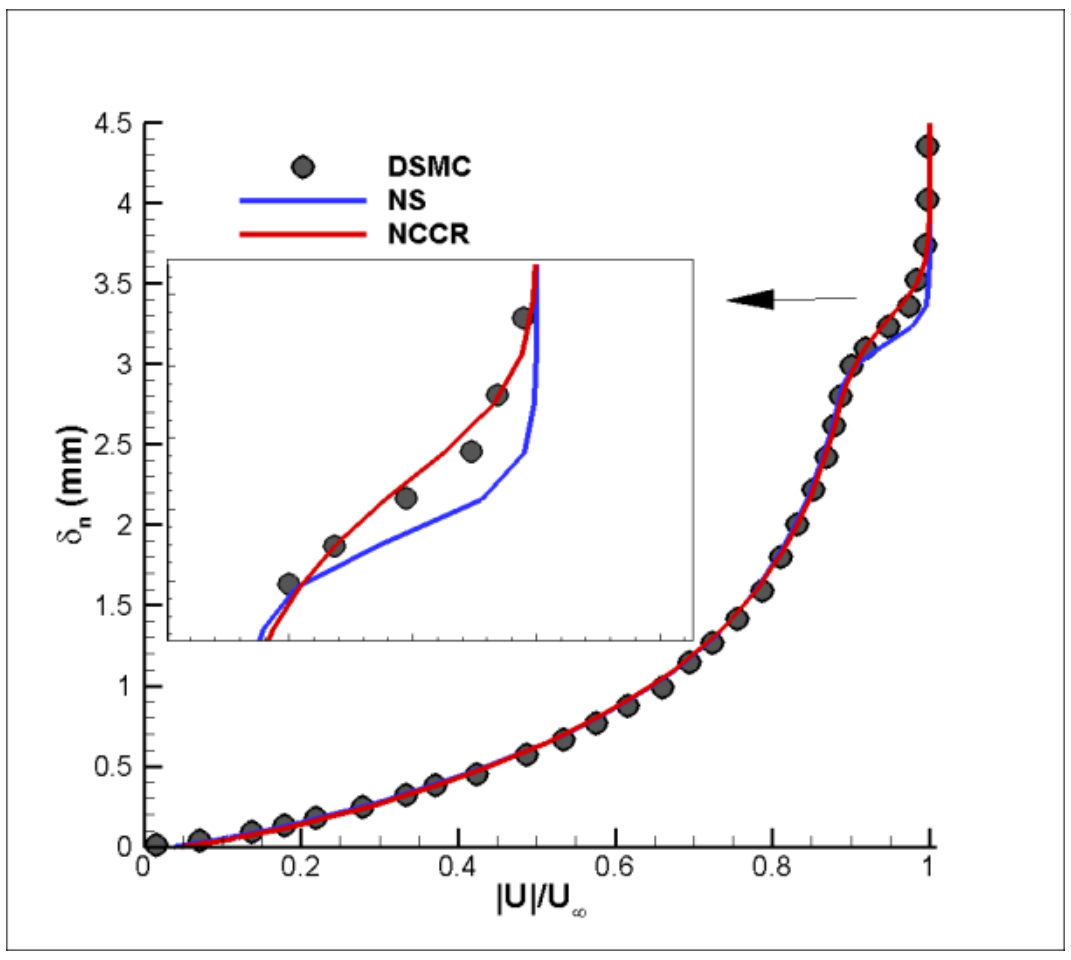

(a) Velocity

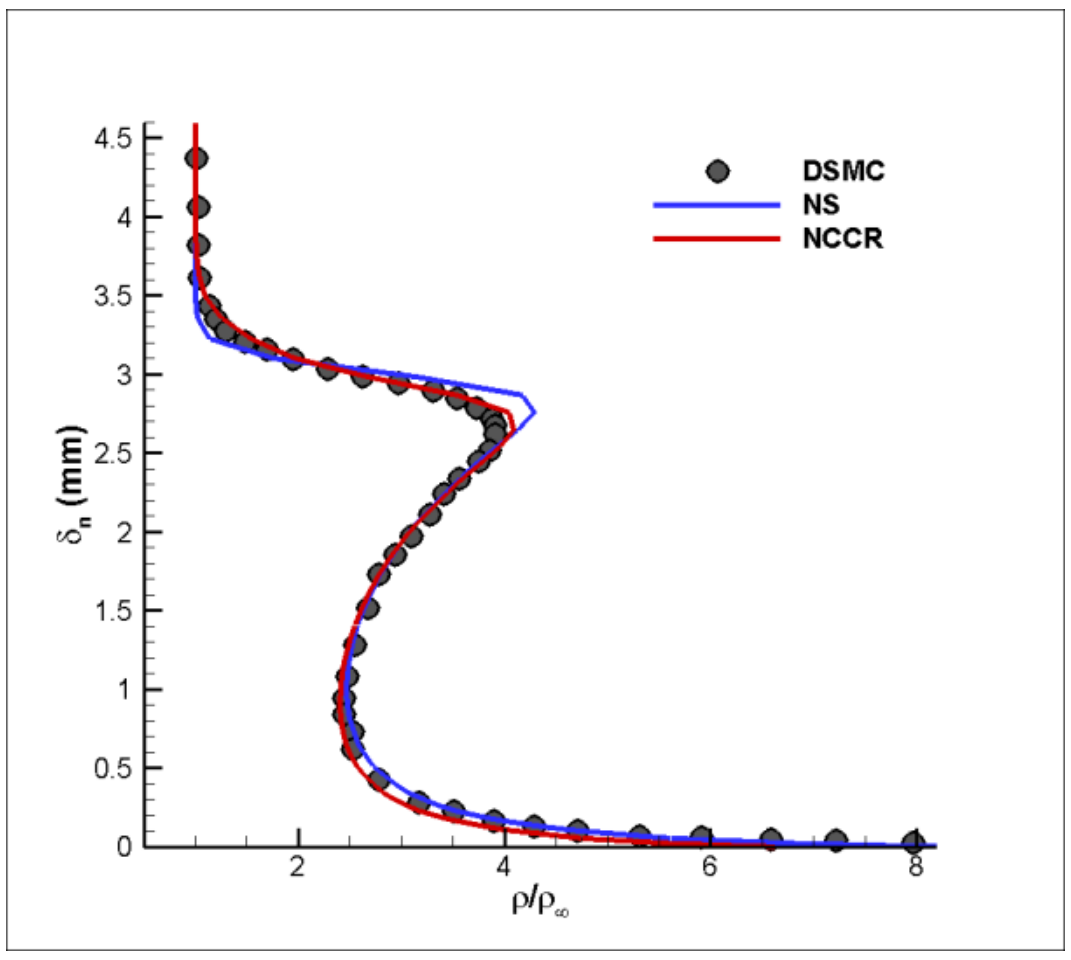

(b) Density

Figure 9 Profiles along the line normal to the cone at $X$= 1.14 cm.

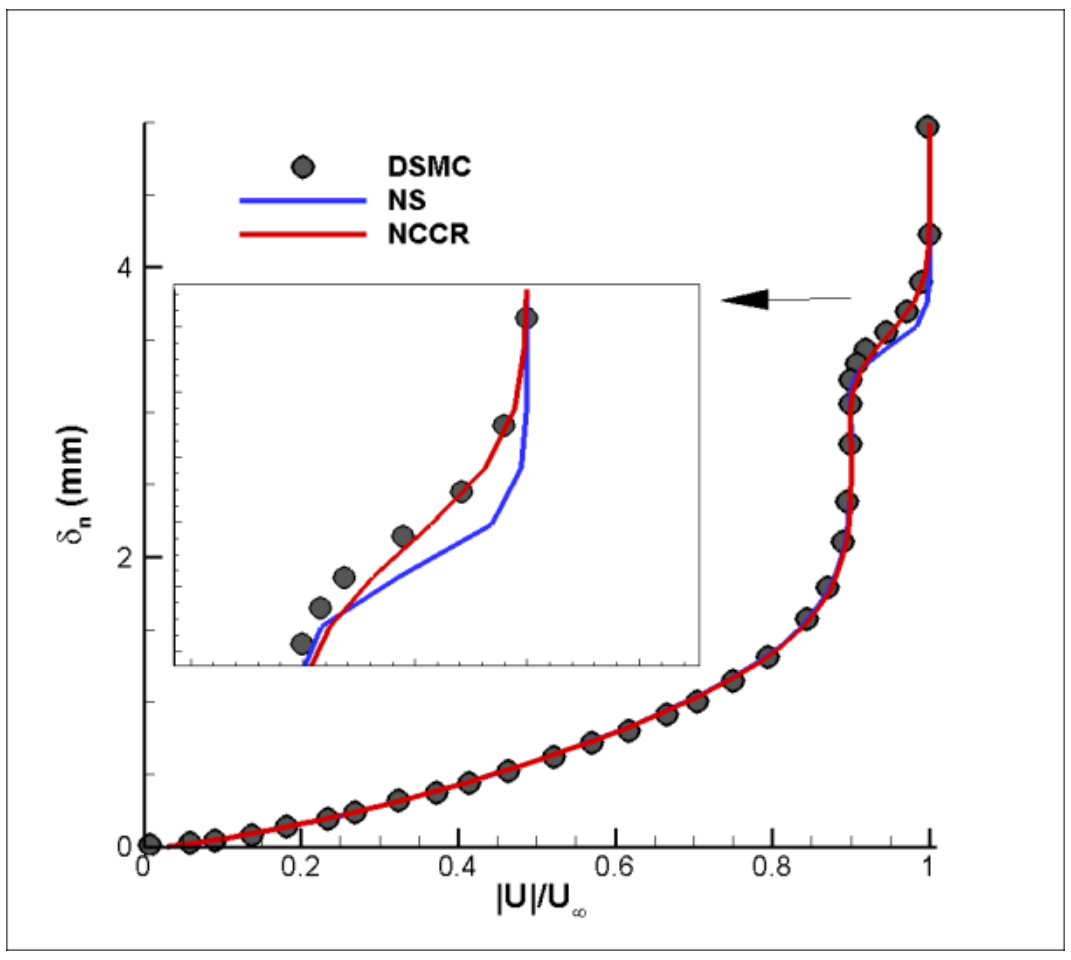

(a) Velocity

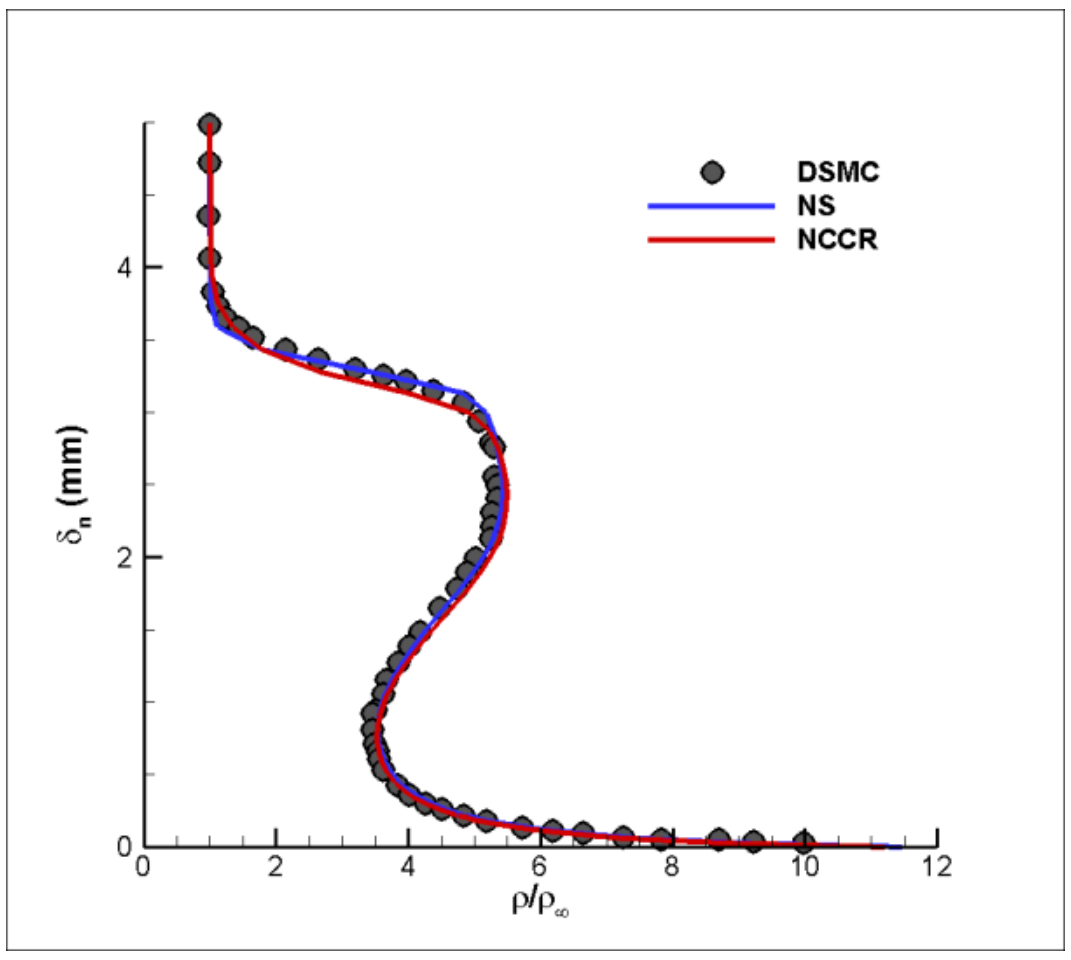

(b) Density

Figure 10 Profiles along the line normal to the cone at $X$= 3.14 cm.

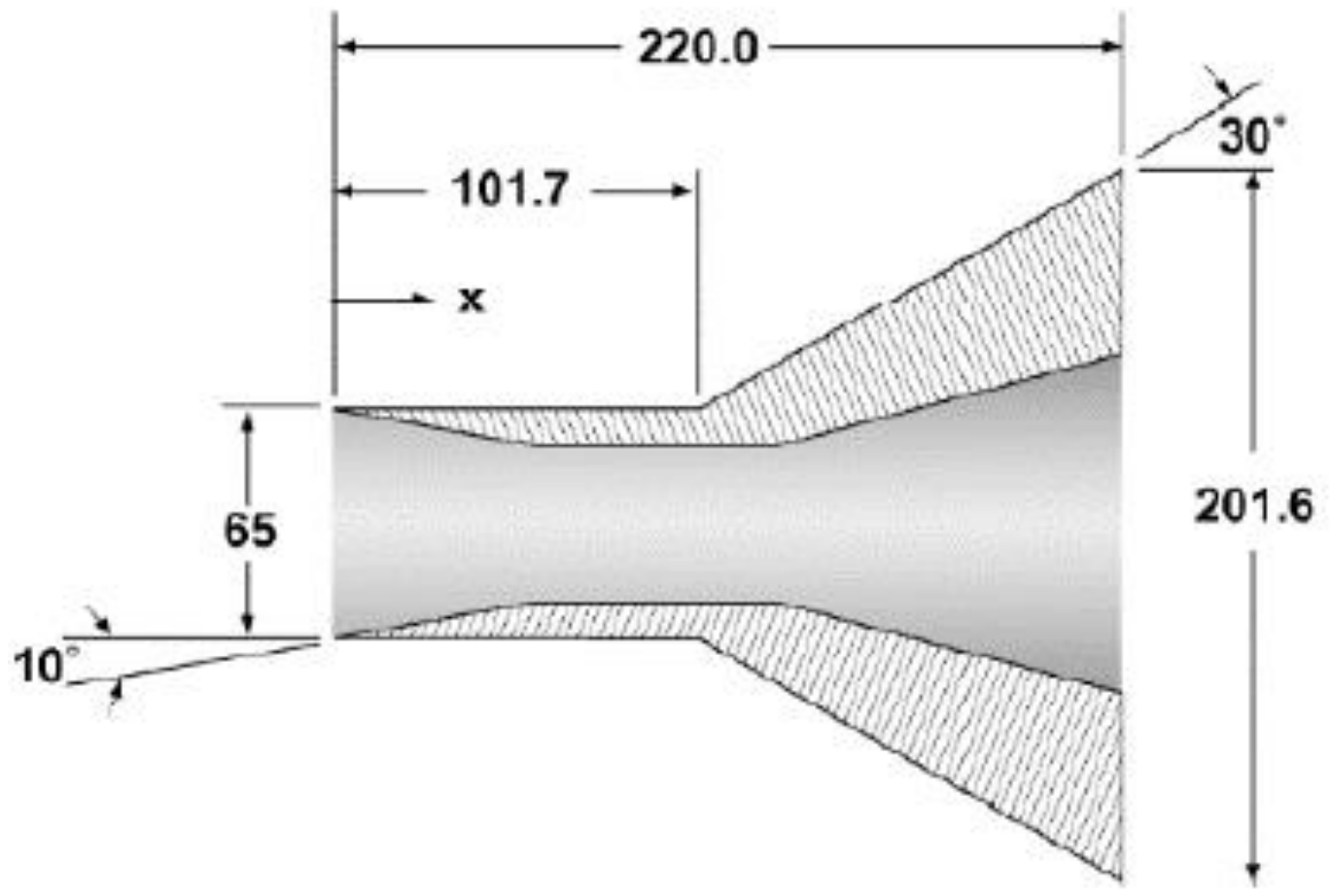

Figure 11 Schematic of hollow cylinder-flare configuration (units in millimeters)[35]

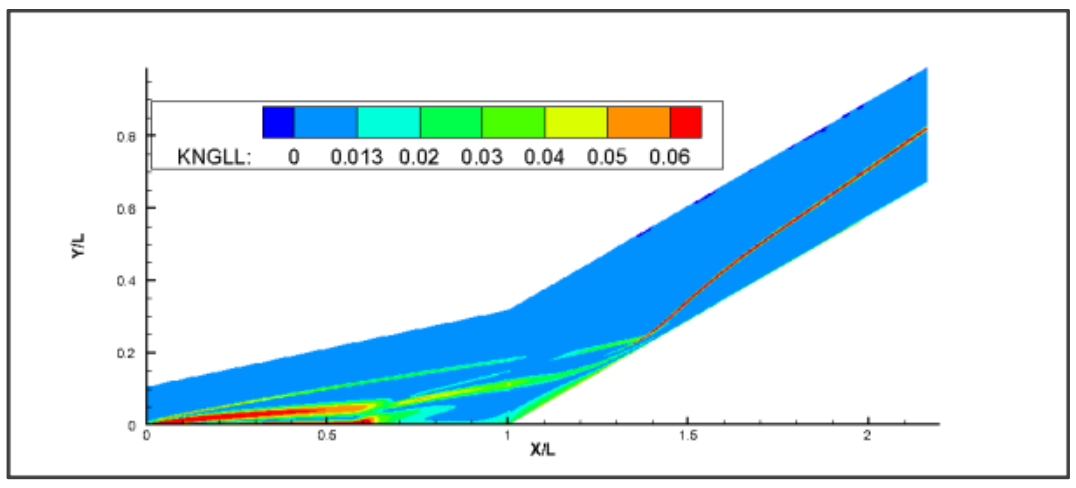

Figure 12 Gradient length local Knudsen contour

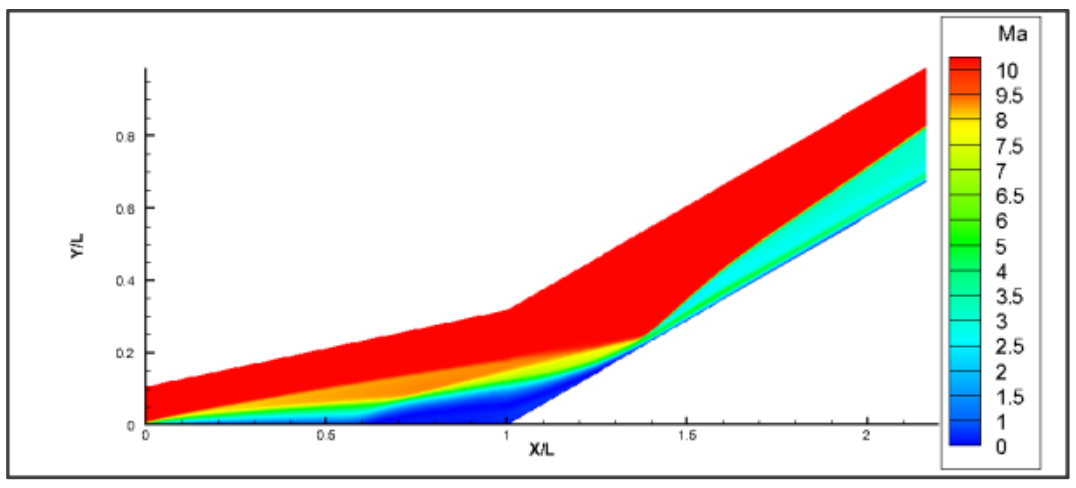

Figure 13 Mach number contour

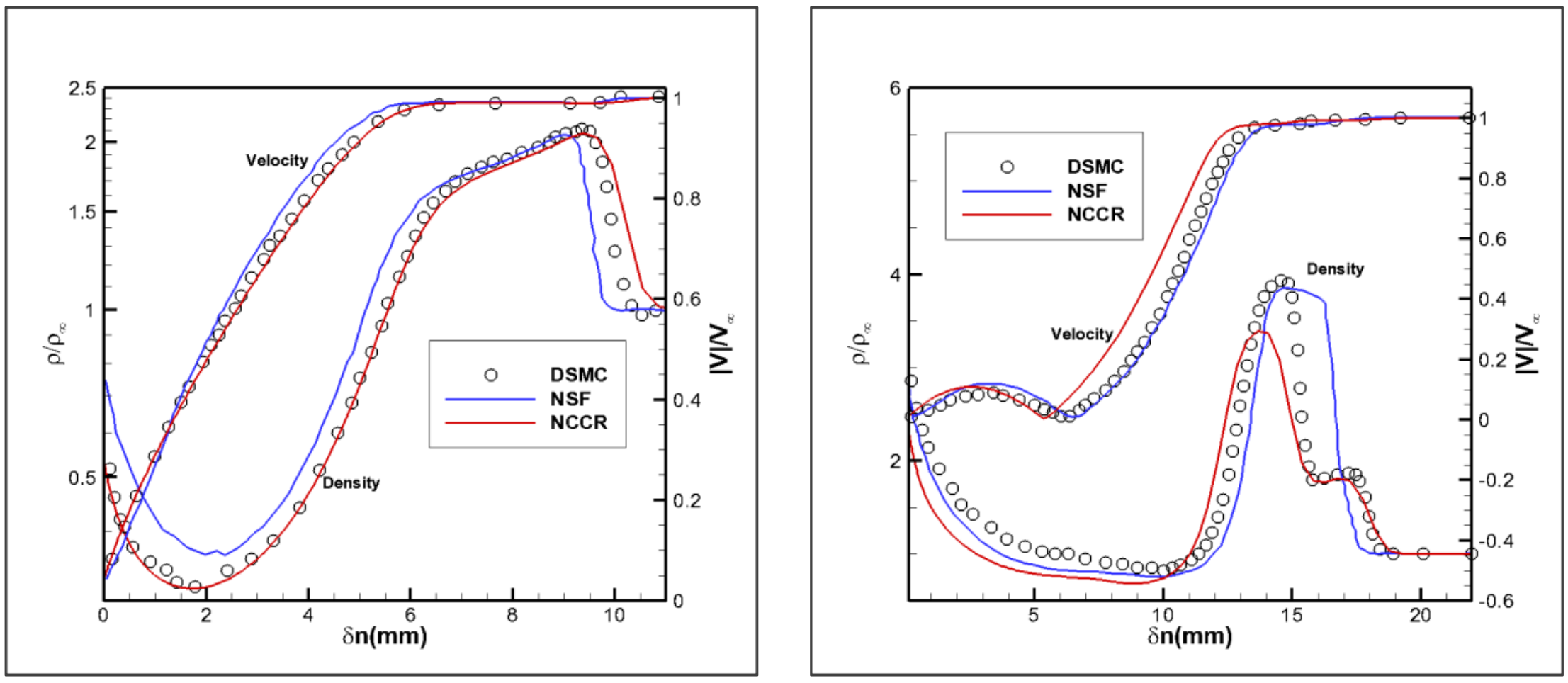

Figure 14 density and velocity profiles along the line normal to the cylinder at x/L=0.5(left) and x/L=1(right)

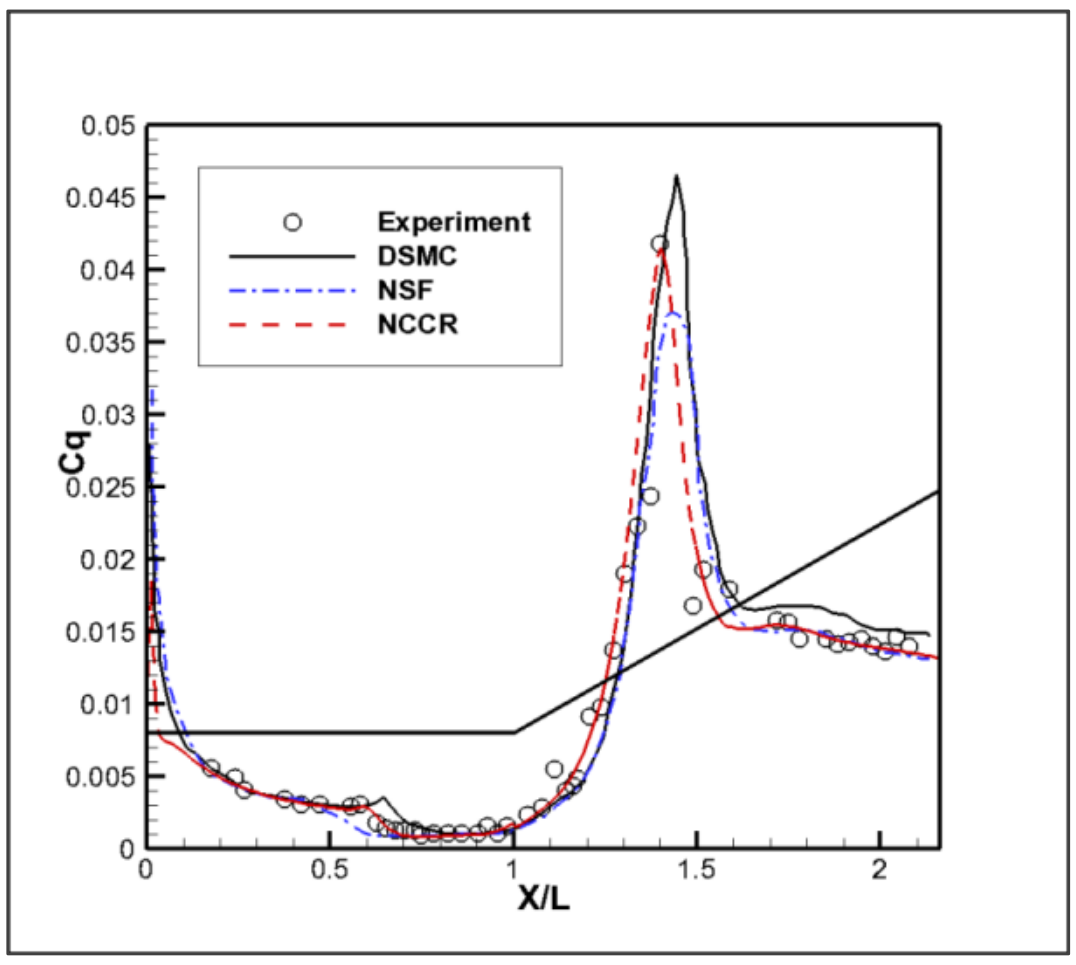

Figure 15 comparison of surface heat flux non-dimensionalized coefficient

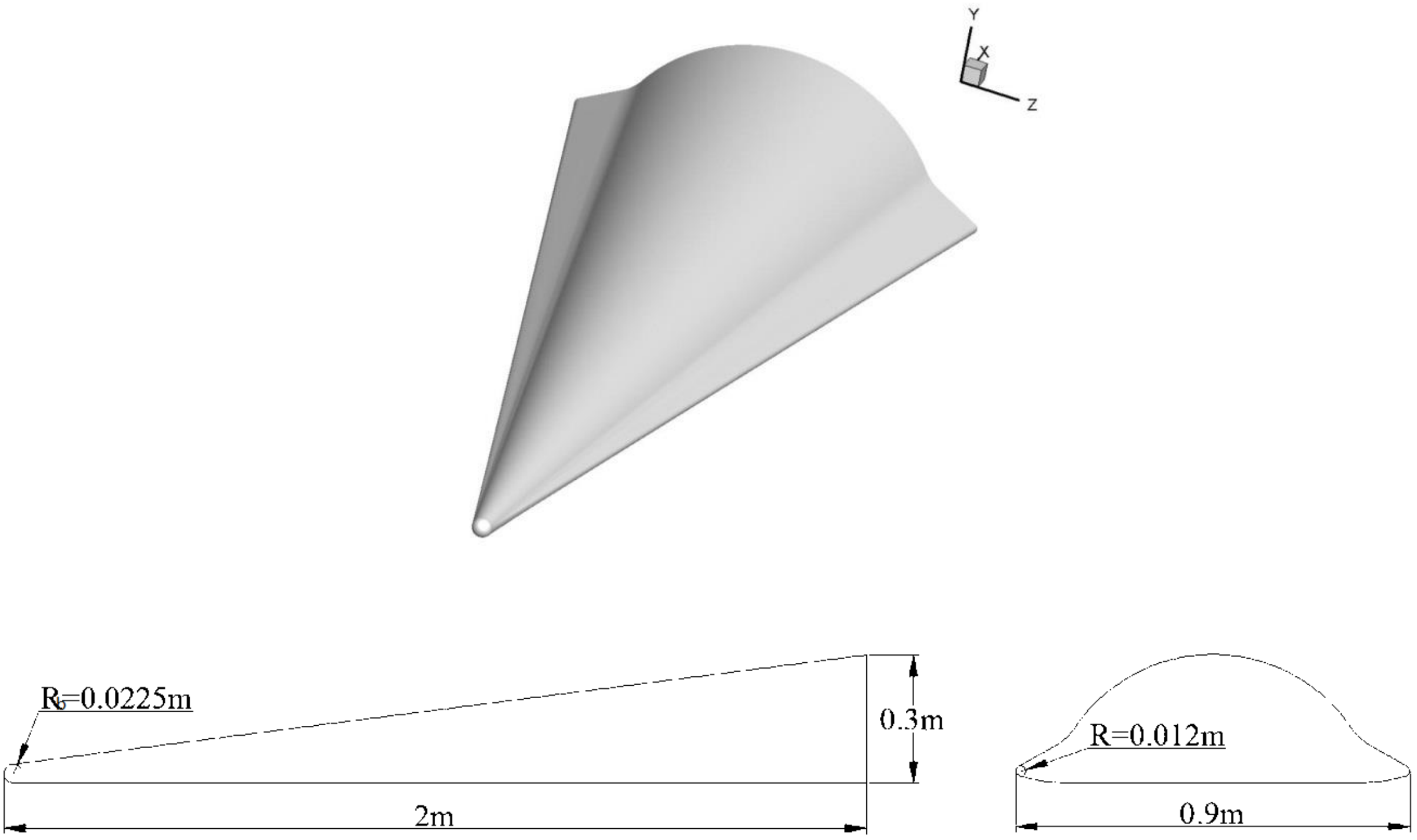

Figure 16: Schematic of the HTV-type vehicle.

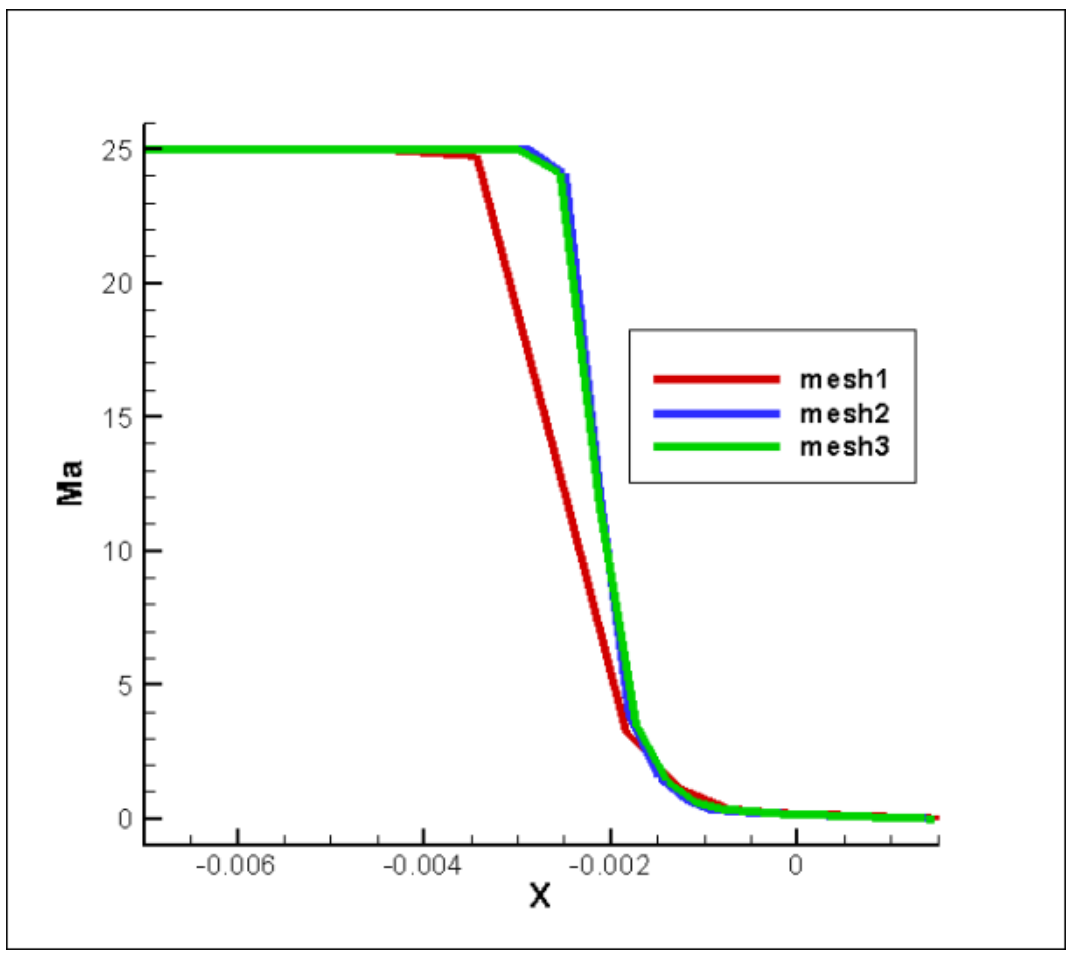

Figure 17: Mach number distributions from three different grid resolutions along the stagnation line ( $Ma = 25$ , $\alpha = 20^o$ , $H = 50\text{km}$ , $T_w = 1000\text{K}$ ).

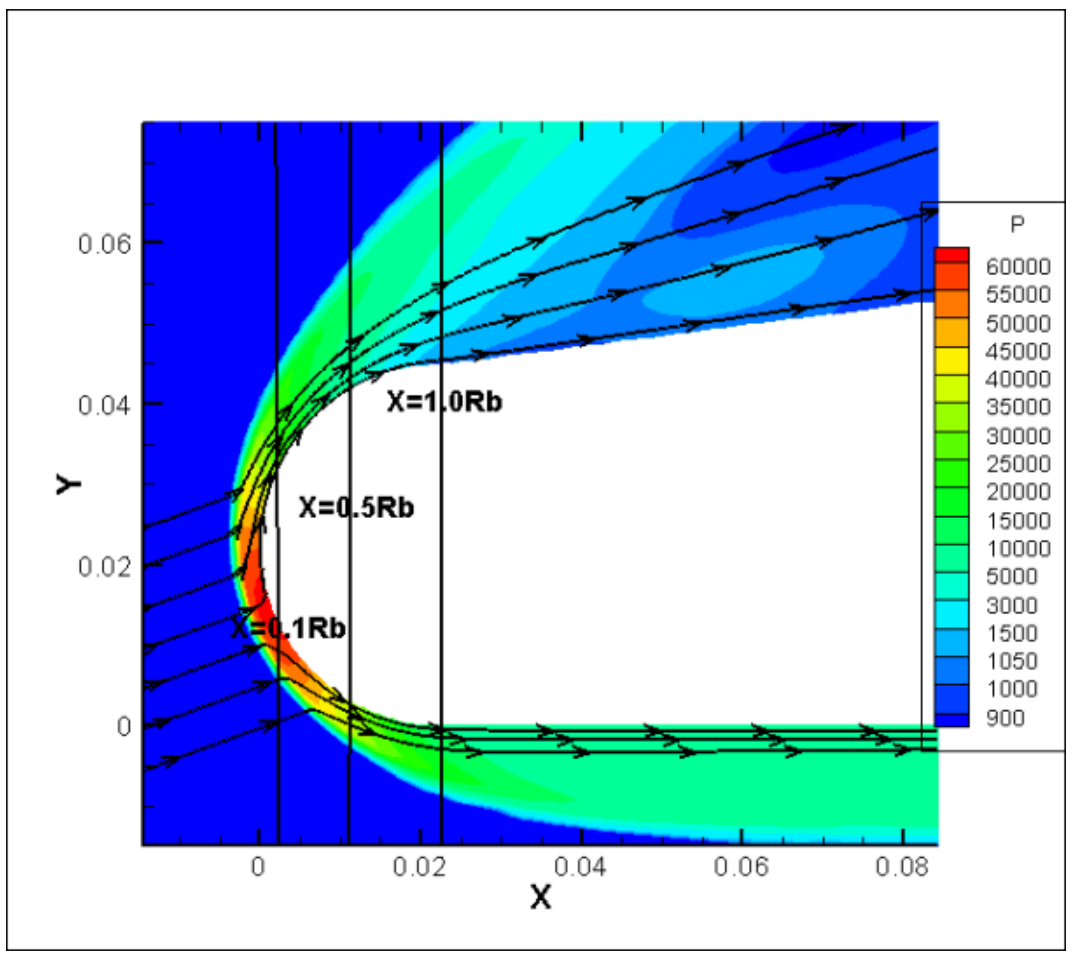

Figure 18: the local contour of pressure and streamline pattern around the sharp leading edge of the HTV-type vehicle computed by NCCR model ( $Ma = 25$ , $\alpha = 20^o$ , $H = 50\text{km}$ , $T_w = 1000\text{K}$ , symmetrical plane $Z = 0$ ).

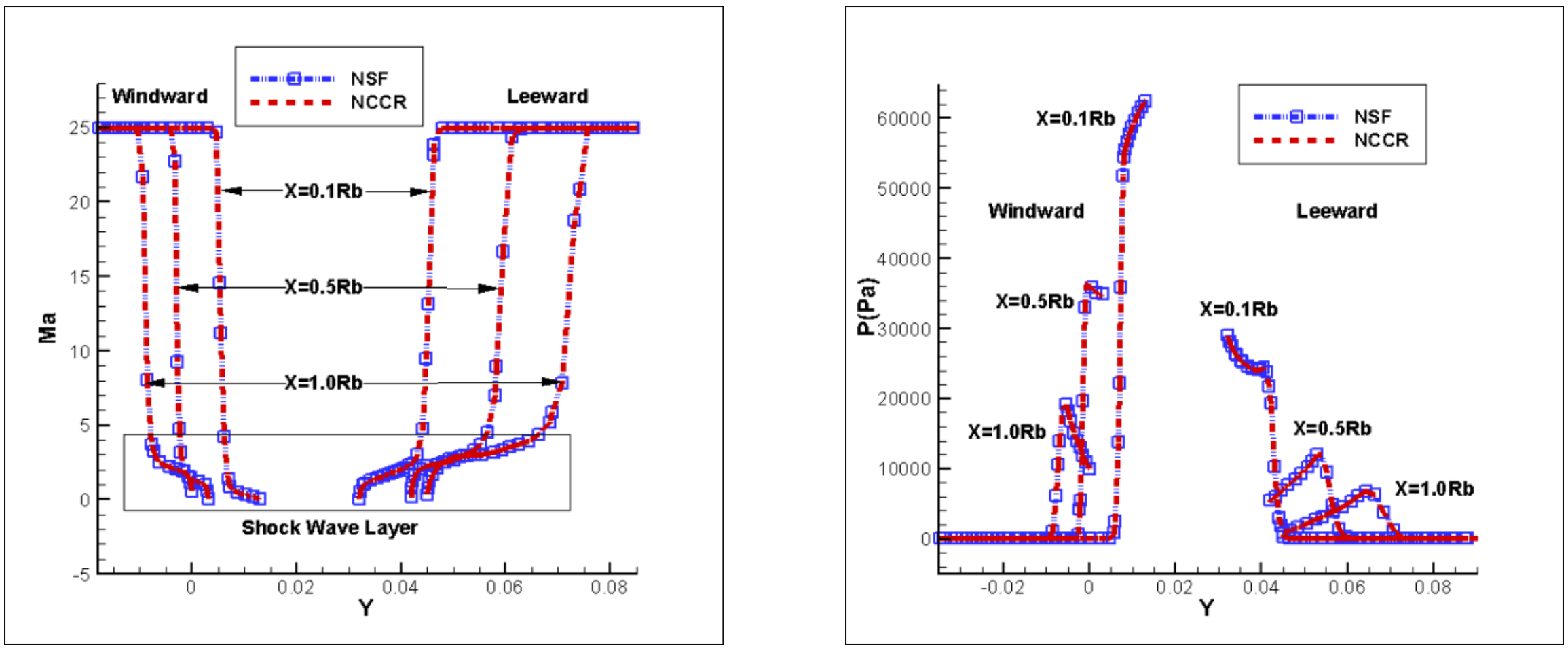

Figure 19: Windward and leeward distributions of Mach number and pressure along y direction on the symmetrical plane ( $Ma = 25$ , $\alpha = 20^o$ , $H = 50\text{km}$ , $T_w = 1000\text{K}$ , symmetrical plane $Z = 0$ ) .

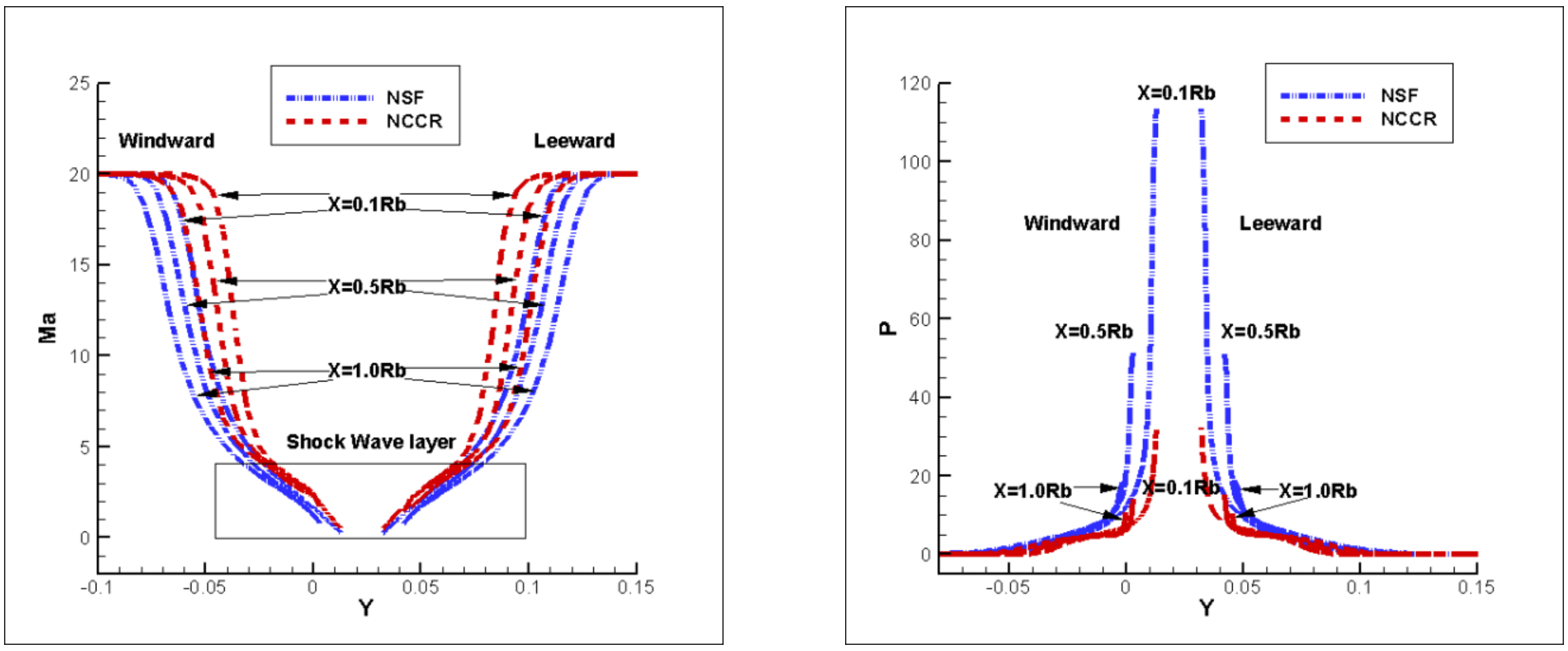

Figure 20: Windward and leeward distributions of Mach number and pressure along y direction on the symmetrical plane ( $Ma = 20$ , $\alpha = 0^o$ , $H = 90\text{km}$ , $T_w = 1000\text{K}$ , symmetrical plane $Z = 0$ ) .

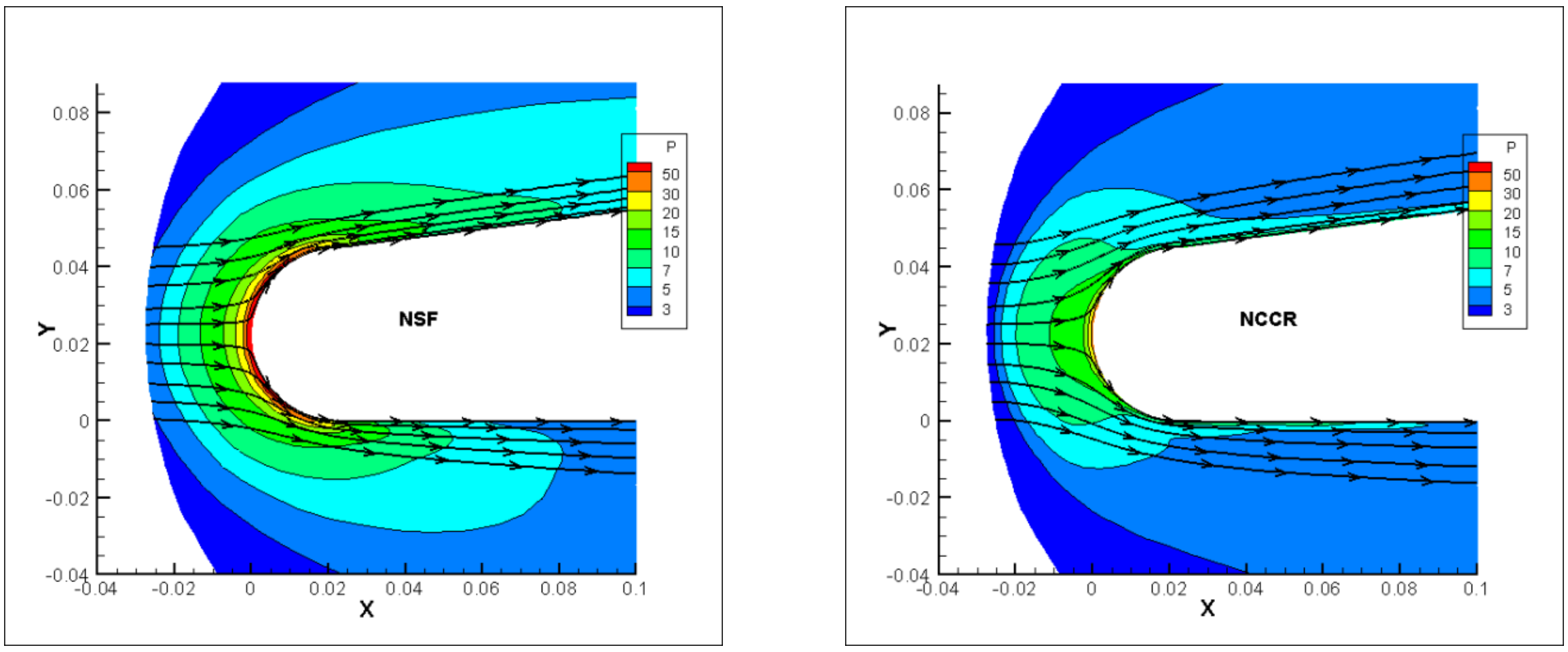

Figure 21: Local contour of pressure and streamline pattern around the sharp leading edge of the HTV-type vehicle computed by NSF and NCCR model ( $Ma = 20$ , $\alpha = 0^o$ , $H = 90\text{km}$ , $T_w = 1000\text{K}$ , symmetrical plane $Z = 0$ ).

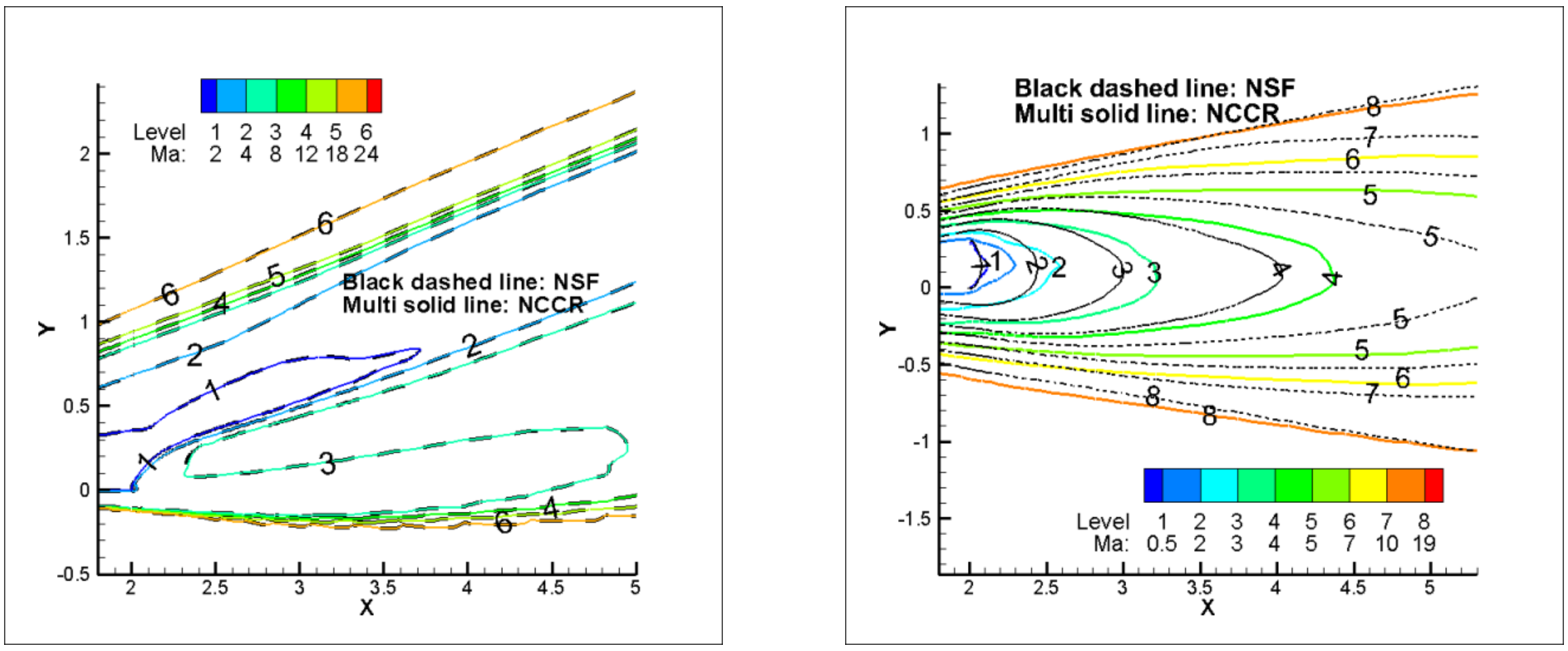

(a) $Ma = 25$ , $\alpha = 20^o$ , $H = 50\text{km}$ , $T_w = 1000\text{K}$ (b) $Ma = 20$ , $\alpha = 0^o$ , $H = 90\text{km}$ , $T_w = 1000\text{K}$

Figure 22: Detailed contour line comparison of after-body flow of Mach number on symmetrical plane $Z = 0$ between NS and NCCR.

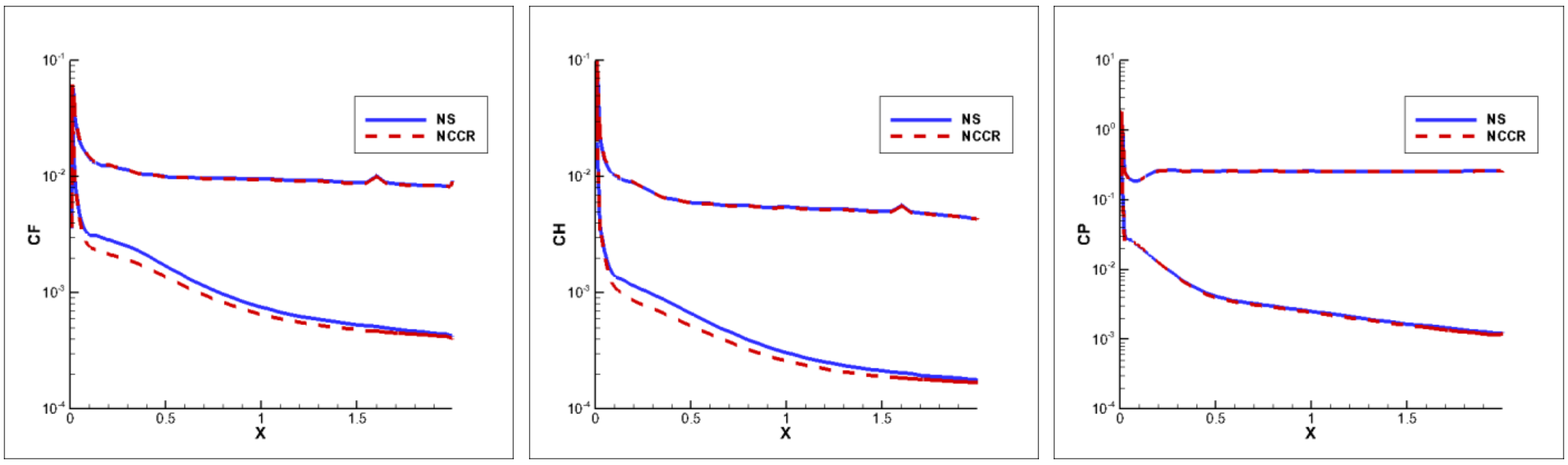

Figure 23 Comparison of surface friction, heating and pressure coefficients ( $Ma = 25$ , $\alpha = 20^o$ , $H = 50\text{km}$ ,

$T_w = 1000\text{K}$ , symmetrical plane $Z = 0$ ).

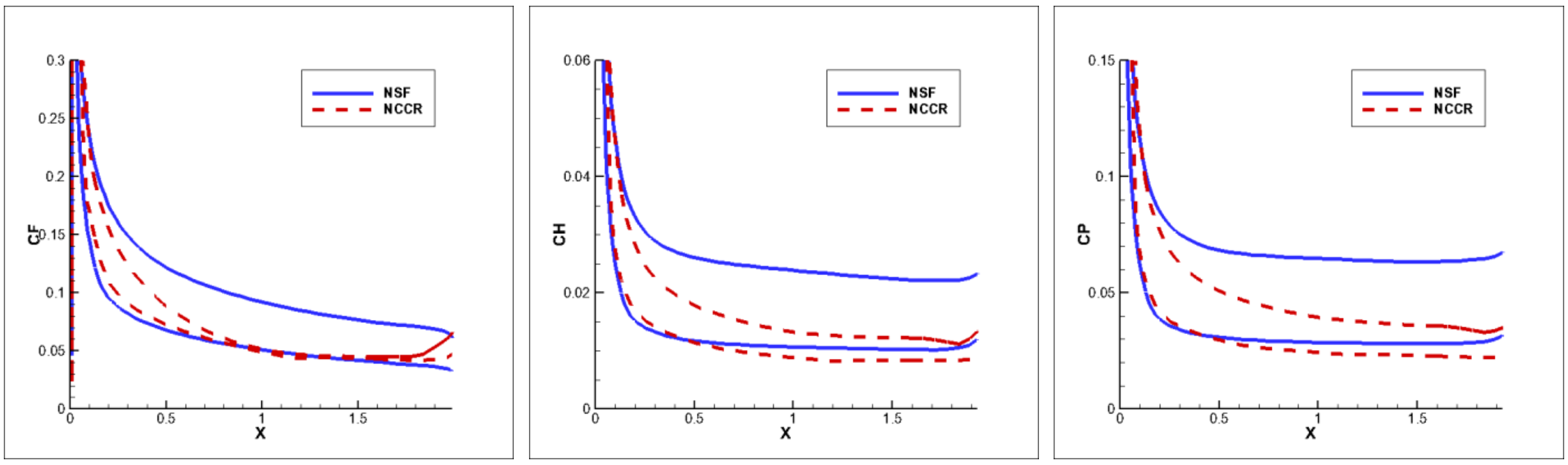

Figure 24: Comparison of surface friction, heating and pressure coefficients ( $Ma = 20$ , $\alpha = 0^o$ , $H = 90\text{km}$ , $T_w = 1000\text{K}$ , symmetrical plane $Z = 0$ ).